\documentclass[aps,prb,twocolumn,superscriptaddress,floatfix]{revtex4-2}
\usepackage[utf8]{inputenc}
\usepackage{natbib}
\usepackage[colorlinks=true,linkcolor=blue,citecolor=blue,urlcolor=blue]{hyperref}
\usepackage{graphicx}
\usepackage{latexsym}
\usepackage{amssymb}
\usepackage{amsmath}
\usepackage{amsfonts}
\usepackage{amsthm}
\usepackage{bm}
\usepackage{floatrow}
\usepackage[caption=false]{subfig}
\usepackage{bbm}
\usepackage{enumitem}
\usepackage{hyperref}
\usepackage{lipsum}
\usepackage{braket}
\usepackage{tikz}
\usepackage{pgfplots}
\usepackage{comment}
\usepackage{ifthen}
\usepackage{ragged2e}
\usepackage{multirow}
\usepackage[export]{adjustbox}
\usetikzlibrary{arrows,decorations.pathreplacing,decorations.markings,arrows.meta,patterns,3d}
\usepackage[normalem]{ulem} 
\usepackage{cancel}
\usepackage{microtype}

\theoremstyle{definition}

\pgfplotsset{compat=1.8}

\begin{document}

\title{Fusion of one-dimensional gapped phases and their domain walls}
\date{\today}

\author{David T. Stephen}
\affiliation{Department of Physics and Center for Theory of Quantum Matter, University of Colorado Boulder, Boulder, Colorado 80309 USA \looseness=-1}
\affiliation{Department of Physics and Institute for Quantum Information and Matter, \mbox{California Institute of Technology, Pasadena, CA, 91125, USA}}
\author{Xie Chen}
\affiliation{Department of Physics and Institute for Quantum Information and Matter, \mbox{California Institute of Technology, Pasadena, CA, 91125, USA}}
\affiliation{Walter Burke Institute for Theoretical Physics, \mbox{California Institute of Technology, Pasadena, CA, 91125, USA}}

\begin{abstract} 
Finite depth quantum circuits provide an equivalence relation between gapped phases. Moreover, there can be nontrivial domain walls either within the same gapped phase or between different gapped phases, whose equivalence relations are given by finite depth quantum circuits in one lower dimension. In this paper, we use such unitary equivalence relations to study the fusion of one-dimensional gapped phases. In particular, we use finite depth circuits to fuse two gapped phases, local unitaries to fuse two domain walls, and a combination of both to fuse gapped phases with domain walls. This provides a concrete illustration of some simple aspects of the `higher-category' structure of gapped defects in a higher-dimensional trivial gapped bulk state.
\end{abstract}

\maketitle


Defects in higher-dimensional quantum phases have recently received a lot of attention~\cite{Roumpedakis2023,Choi2022,Kaidi2023,Bhardwaj2018,Bhardwaj2023,Bhardwaj2023b,Aasen2016,Chang2019, Thorngren2019fusion,Johnson-Freyd2022,Ji2022,Kong2020algebraic}. Some of them satisfy a non-invertible fusion rule \cite{Roumpedakis2023,Choi2022,Kaidi2023,Bhardwaj2018,Aasen2016,Bhardwaj2023b,Ji2022} and generalize the usual notion of symmetry transformations which are unitary / anti-unitary and invertible. Gapped defects in non-trivial topological phases have been proposed to have a category or higher category structure \cite{Thorngren2019fusion,Johnson-Freyd2022,Kong2020algebraic,Bhardwaj2023}. In particular, a 2-category structure has been proposed to describe one-dimensional (1D) defects where the objects of the 2-category are 1D gapped defects and the morphisms between the objects are the zero-dimensional (0D) domain walls either within the same defect or between different defects. 

Gapped defects and excitations in a topologically trivial phase should also have a higher category structure. 1D gapped defects in a higher-dimensional trivial bulk state are simply 1D gapped phases and their morphisms are domain walls either within the same phase or between different phases. In this paper, we study some simple aspects of the 2-category structure of 1D gapped phases and their 0D domain walls. On the one hand, this provides a concrete and simple physics context to illustrate some aspects of the higher-category structure discussed in the math literature. On the other hand, understanding defects in trivial phases is a pre-requisite for the proper understanding of defects in topologically nontrivial phases as the 1D phases exist as decoupled defects in topological phases and more importantly, they show up as coefficients in the fusion of non-decoupled defects (see for example Ref.~\onlinecite{Roumpedakis2023,Kong2020defects}. Discussions of the 2-category structure of 1D gapped phases can be found in Ref.~\onlinecite{Bhardwaj2023b,Kong2022b}.

Our analysis uses unitary quantum circuits and is based on the following rule:

\textit{The equivalence relation between 1D gapped defects is given by 1D finite depth circuits while the equivalence relation between 0D domain walls is given by local unitary operations.}

\begin{figure}[ht]
    \centering
    \includegraphics[scale=0.50]{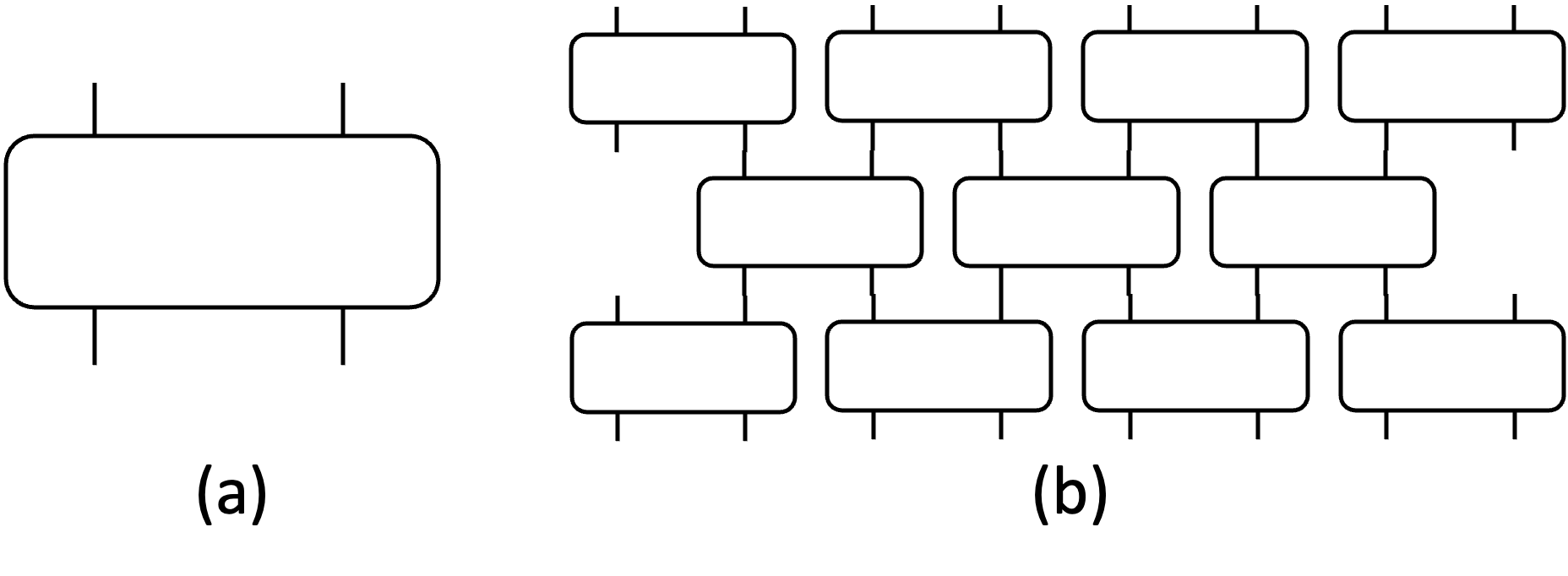}
    \caption{(a) 0D local unitary operations and (b) 1D finite depth quantum circuit.}
    \label{fig:LU}
\end{figure}

Fig.~\ref{fig:LU} is an illustration of (a) 0D local unitary operations and (b) 1D finite-depth circuits. These equivalence relations are natural extensions of the definition for super-selection sectors of quasi-particle excitations \cite{Kitaev2006} and the finite depth circuit equivalence between 1D gapped phases \cite{Chen2010}. We use these equivalence relations to study the equivalence classes of 1D gapped phases and their domain walls. Moreover, we derive fusion rules between 1D gapped phases, between their domain walls, and between 1D gapped phases with domain walls.  

Some of the interesting results we get include:
\begin{enumerate}
\item Fusion of a symmetry breaking phase with a symmetric phase results in a symmetry breaking phase.
\item Nontrivial domain walls on symmetry breaking phases are `flux' domain walls between order parameters, while nontrivial domain walls on symmetric phases are symmetry charges. 
\item Fusion of two symmetry breaking phases results in a symmetry breaking phase with a coefficient that is a 1D system with degeneracy not protected by the symmetry.
\item When one of the symmetry breaking phases contains a domain wall, the fusion coefficient contains a domain wall.
\end{enumerate}

The paper is structured as follows. In Section~\ref{sec:phases}, we discuss possible 1D gapped phases together with their domain walls, especially in the presence of a symmetry. In Section~\ref{sec:fusion}, we use finite depth quantum circuits to fuse 1D gapped phases without domain walls. In Section~\ref{sec:fusion_dw}, we use 0D local unitaries to fuse domain walls. In Section~\ref{sec:fusion+}, we discuss how to use a combination of 0D local unitaries and 1D finite depth circuits to fuse 1D gapped phases with domain walls. The situation in higher dimensions is much more complicated with the appearance of nontirival topological order, but we discuss some simple cases in Section~\ref{sec:2D}. 

Throughout the discussion, we will make frequent use of operators of the form $ e^{-\frac{i\pi}{4} \mathcal O}$. We define short hand notation $R(\mathcal O) \equiv  e^{-\frac{i\pi}{4} \mathcal O}$, which has the property that for Pauli operators $P$ and $Q$,
\begin{align}
    R(Q) P R(Q)^\dagger = \begin{cases}
    P ;& [P,Q]=0\\
    iPQ; & \{P,Q \}=0
    \end{cases}
\end{align}

\section{1D gapped phases and domain walls}
\label{sec:phases}

1D gapped phases have been completely classified with or without global symmetry \cite{Chen2011classification,Schuch2011,Fidkowski2011}. Without global symmetry, there are no non-trivial phases, whereas with global symmetry, there are different possibilities. First the global symmetry $G$ can be spontaneously broken if it is finite, giving rise to a symmetry broken (SB) phase with nontrivial ground state degeneracy. When the symmetry is not broken, there is the possibility of having a symmetry protected topological (SPT) phase classified by the cocycle group $H^2(G,U(1))$. The bulk of the SPT phases are gapped and non-degenerate while the edge carries non-trivial degeneracy. Finally, there is the possibility of partial symmetry breaking from $G$ to a subgroup $H$ combined with symmetry protected topological order of $H$ classified by $H^2(H,U(1))$. In this work, we will focus our attention on cases where the symmetry is either preserved or completely broken. We make some comments on the case of partial symmetry breaking in Sec.~\ref{sec:partialssb}.

It is known that finite depth quantum circuits connect ground states within the same gapped 1D phase \cite{Chen2010}. For a system with global symmetry $G$, the circuit is symmetric in the sense that each local gate is invariant under the symmetry. Ground states of different gapped phases on the other hand, cannot be mapped into each other through finite depth circuit \cite{Chen2010}. Instead, a sequential quantum circuit is needed \cite{Huang2015,Chen2024sequential}.

What kinds of 0D domain walls exist within each 1D gapped phase? A domain wall is a local excitation on top of the ground state. On the two sides of the domain wall, the reduced density matrix looks exactly like that in the ground state while on the domain wall the reduced density matrix can be different. A domain wall is nontrivial if the reduced density matrix around the domain wall is different from that of the ground state and if it cannot be created with (symmetric) local unitaries at the location of the domain wall. Otherwise, the domain wall is trivial. Two domain walls are equivalent to each other if they can be mapped to each other through a (symmetric) local unitary.

Applying this rule, we can find domain walls for different gapped phases.
\begin{enumerate}
\item In a symmetric phase (trivial or non-trivial SPT), a non-trivial domain wall is an isolated charge. 
\item In the symmetry-breaking phase, a non-trivial domain wall is a domain wall between different values of the order parameter, which we refer to as a flux. 
\end{enumerate}

Fig.~\ref{fig:phases} gives the graphical representation of (a) a symmetric phase with a charged domain wall, and (b) a symmetry-breaking phase with a `flux' domain wall. In general, a charged domain wall can be generated from the ground state at some site $k$ by applying a local charged operator around $k$, while a flux domain wall can be generated by applying the broken symmetry to all sites $i>k$. Neither of these correspond to local symmetric operators, and hence they generate non-trivial domain walls.

Prototypical examples of these two cases can be given by the transverse field Ising model. In the symmetric phase with the Hamiltonian 
\begin{equation}
H = -\sum_i X_i\, ,
\label{eq:SYH}
\end{equation}
and ground state wavefunction
\begin{equation}
|...+++++...\rangle\, ,
\label{eq:SYpsi}
\end{equation}
a charged domain wall at site $k$ corresponds to changing the sign of the Hamiltonian term at site $k$ to $+ X_k$
\begin{equation}
H = -\sum_{i\neq k} X_i + X_k\, .
\end{equation}
The ground state wave function takes the form
\begin{equation}
|...++-++...\rangle\, .
\label{eq:SYpsi_dw}
\end{equation}
This charge can be generated by acting with a charged operator $Z$ at site $k$.

In the symmetry breaking phase with the Hamiltonian 
\begin{equation}
H=-\sum_i Z_iZ_{i+1}
\label{eq:SBH}
\end{equation}
and symmetrized ground state,
\begin{equation}
|...0000...\rangle + |...1111...\rangle
\label{eq:SBpsi}
\end{equation}
a flux domain wall between sites $k$ and $k+1$ corresponds to changing the sign of the corresponding Hamiltonian term to $+Z_kZ_{k+1}$,
\begin{equation}
H = -\sum_{i\neq k} Z_iZ_{i+1} + Z_kZ_{k+1}\, .
\end{equation}
The (symmetrized) ground state wave function takes the form
\begin{equation}
|...0011...\rangle + |...1100...\rangle\, .
\label{eq:SBpsi_dw}
\end{equation}
The flux domain wall can be generated between sites $k$ and $k+1$ by acting with the broken symmetry on all sites $i>k$.

A prototypical nontrivial SPT phases is given by the cluster state with $Z_2\times Z_2$ symmetry \cite{Raussendorf2001a,Son2012}. The system contains two sets of qubits, one on integer lattice sites and one on half integer ones, each transforming under a $Z_2$ symmetry $\prod_i X_i$ and $\prod_{i} X_{i+1/2}$. The Hamiltonian of the cluster state is 
\begin{equation}
H = \sum_i - Z_{i-\frac12}X_iZ_{i+\frac12} - Z_iX_{i+\frac12}Z_{i+1}
\label{eq:SPTH}
\end{equation}
and the ground state wave function is given by
\begin{equation}
\prod_i CZ_{i-\frac12,i}CZ_{i,i+\frac12} |...++++++...\rangle\, .
\label{eq:SPTpsi}
\end{equation}
{Because the symmetry is now $Z_2\times Z_2$, there are two different symmetry charges.
A domain wall charged under the first $Z_2$ can be generated at site $k$ by acting with $Z$, corresponding to changing the sign of the Hamiltonian term at site $k$ to $+ Z_{k-1/2}X_kZ_{k+1/2}$. 
\begin{equation}
\begin{aligned}
H = & \sum_i - Z_{i-\frac12}X_iZ_{i+\frac12} - Z_iX_{i+\frac12}Z_{i+1} \\
& + 2Z_{k-\frac12}X_kZ_{k+\frac12}\, .
\end{aligned}
\end{equation}
The ground state changes to
\begin{equation}
\prod_i CZ_{i-\frac12,i}CZ_{i,i+\frac12} |...++-+++...\rangle\, .
\label{eq:SPTpsi_dw}
\end{equation}
We can similarly generate a domain wall charged under the second $Z_2$ by acting with $Z$ on a site $k+1/2$.
We can also try to generate flux domain walls in the cluster state by, e.g., acting with one of the $Z_2$ symmetries on all half integer sites to the right of site $k$. But, this will have the same effect as acting with $Z$ on site $k$, i.e. inserting a charged domain wall, so this does not generate a new kind of domain wall.

There are no other nontrivial class of domain walls: a symmetry flux in a symmetric state gives rise to either a trivial domain wall in a trivial symmetric state or a charged domain wall in a symmetry protected topological state, as seen above. On the other hand, a charged domain wall in the symmetry-breaking phase disappears into the bulk and is not locally detectable and is hence a trivial domain wall. We give general arguments for these claims using matrix product states in Appendix \ref{appen:MPS}.

\begin{figure}[ht]
    \centering
    \includegraphics[scale=0.65]{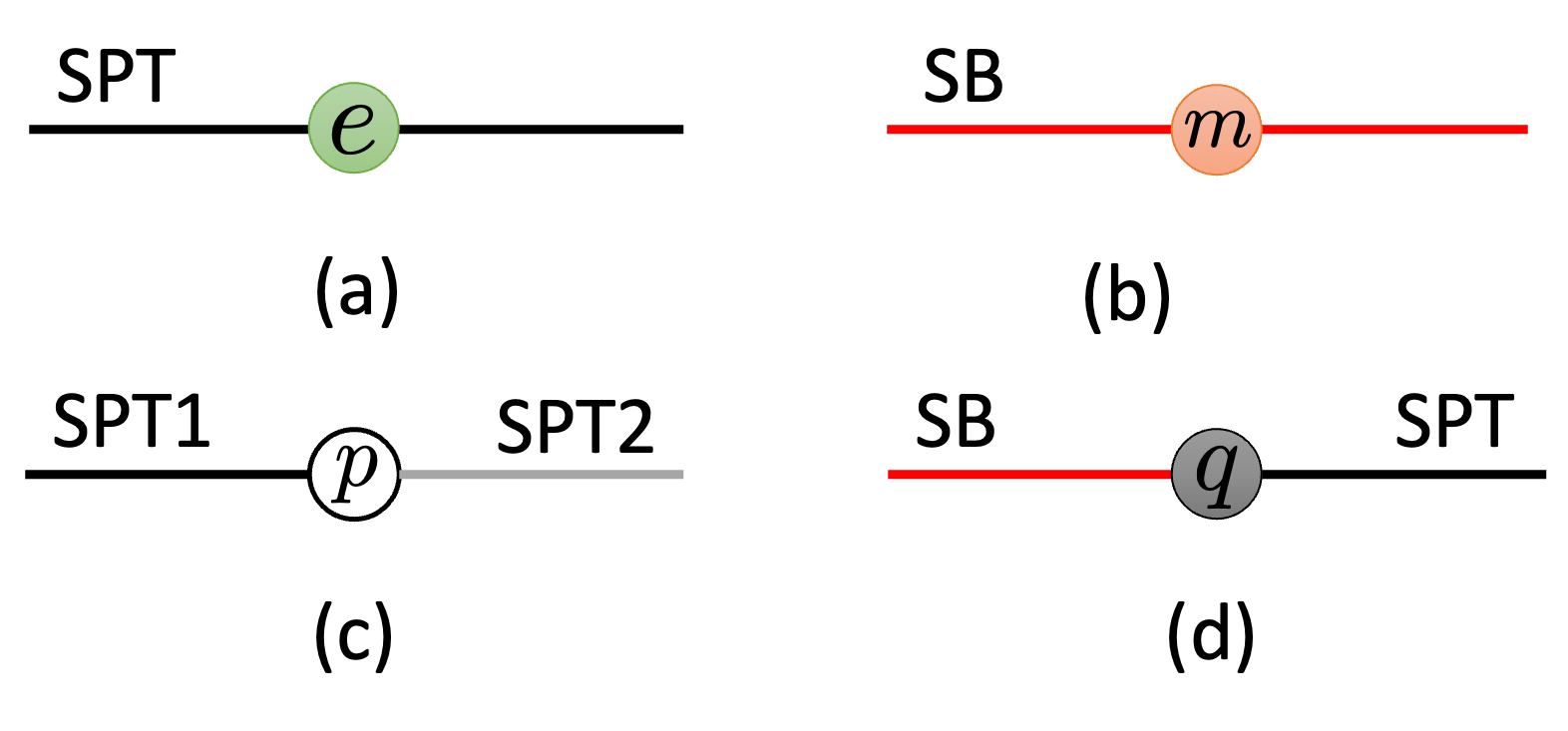}
    \caption{Gapped 1D phases with domain wall: (a) A symmetric phase with a charged domain wall labeled by $e$. (b) A symmetry breaking phase with a flux domain wall labeled by $m$. (c) A degenerate domain wall between different SPT phases. (d) A non-degenerate domain wall between symmetry breaking and SPT phases.}
    \label{fig:phases}
\end{figure}


If the system contains fermions, the Majorana chain gives a nontrivial phase \cite{Kitaev2001}. It is similar to the bosonic SPT phases in that a fermion parity symmetry flux is equivalent to a symmetry charge. Therefore, there is one type of nontrivial domain wall on the Majorana chain in the form of a fermion. 

There are also domain walls between different phases. Between different SPT phases, there is a degenerate projective edge mode denoted by the circle in Fig.~\ref{fig:phases} (c). We denote such domain walls as $p$. For example, between the  cluster state at $i\leq 0$ and the trivial phase at $i>0$ with the Hamiltonian
\begin{equation}
\begin{aligned}
H = &\sum_{i<0} - Z_{i-\frac12}X_iZ_{i+\frac12} - Z_iX_{i+\frac12}Z_{i+1} \\
+&\sum_{i>0} -X_{i-\frac12} - X_i
\end{aligned}
\end{equation} 
the 2-fold degenerate edge mode is given by the anticommuting operators $Z_{-1/2}X_0$ and $Z_0$ which both commute with $H$ and are each charged under one of the $Z_2$ symmetries. The wavefunction across the domain wall cannot be invariant under the full $Z_2\times Z_2$ group. One possible form that is invariant under $\prod_i X_{i+1/2}$ but not under $\prod_i X_{i}$ is given by 
\begin{equation}
CZ_{-\frac12,0} \prod_{i<0} CZ_{i-\frac12,i}CZ_{i,i+\frac12}|...+++...\rangle\, .
\end{equation}

In fermion systems, the domain wall between the Majorana chain and a trivial chain contains a Majorana zero mode. 

When an SPT phase is connected to a symmetry-breaking phase, the edge mode of the SPT phase can be coupled to the order parameters in the symmetry-breaking phase, thereby hiding the degeneracy of the projective edge mode behind the bulk degeneracy in the symmetry-breaking phase. This is shown in Fig.~\ref{fig:phases} (d) with a solid point and is denoted by $q$. If the SPT phase is on the left-hand side and the symmetry-breaking phase on the right-hand side, we denote the domain wall as $\bar{q}$. For example, between the cluster state at $i\leq 0$ and the symmetry breaking phase at $i>0$ with the Hamiltonian 
\begin{equation}
\begin{aligned}
H = &\sum_{i<0} - Z_{i-\frac12}X_iZ_{i+\frac12} - Z_iX_{i+\frac12}Z_{i+1}
\\+&\sum_{i>0} -Z_iZ_{i+1} - Z_{i-\frac12}Z_{i+\frac12}
\end{aligned}
\end{equation}
a symmetric term $Z_{-1/2}X_0Z_{1/2}$ can be added at the domain wall to couple the edge mode (acted upon by $Z_{-1/2}X_0$) to the order parameter $Z_{1/2}$ and `merge' the degeneracy. This reduces the eight-fold degeneracy (the twofold degeneracy of the SPT edge combined with the four-fold degeneracy of the symmetry-breaking phase) to a net four-fold degeneracy. The four degenerate states across the domain wall are then given by
\begin{align}
&CZ_{-\frac12,0} \prod_{i<0} CZ_{i-\frac12,i}CZ_{i,i+\frac12} \times \\
&|...++00...\rangle, \ |...++01...\rangle, \  |...+-10...\rangle, \ |...+-11...\rangle\, . \nonumber
\end{align}
Since the degenerate bulk states of the symmetry breaking phase combine into different integer charged states under the global symmetry, it cannot screen the projective mode on the SPT edge. The degeneracy associated with the projectiveness of the SPT edge will resurface when we study the fusion of such domain walls in the section~\ref{sec:fusion_dw}. 

There are of course other ways of connecting the symmetry breaking side and SPT side, for example with Hamiltonian terms $-Z_{1/2}X_0Z_{1/2}$, $-Z_0Z_1$, $+Z_0Z_1$. They can all be mapped to each other through symmetric local unitary transformations, hence confirming that there is only one types of domain wall between an SB and an SPT state. To map between $+Z_{1/2}X_0Z_{1/2}$ and $-Z_{1/2}X_0Z_{1/2}$, we can use $Z_0Z_1$; to map between $-Z_0Z_1$ and $+Z_0Z_1$, we can use $Z_{1/2}X_0Z_{1/2}$; to map between $Z_{1/2}X_0Z_{1/2}$ and $Z_0Z_1$, we can use $R(Z_0Z_1)R(Z_{1/2}X_0Z_{1/2})$. 


\section{Fusion of gapped phases}
\label{sec:fusion}

We can fuse gapped phases by stacking them on top of each other and applying a finite depth circuit. When the system has a global symmetry, each local gate in the circuit needs to be symmetric. In the following subsections, we are going to explicitly construct the circuit that realizes the fusion for different pairs of phases. In particular, we are going to choose a standard form for each phase and use finite depth circuits to map the stack of two phases into the standard form of a third. The result of the fusion is shown in Fig.~\ref{fig:fusion}.
\begin{enumerate}[label=\alph*)]
\item The fusion of a symmetry-breaking with a symmetric phase results in a symmetry breaking phase. This holds no matter what SPT order the symmetric phase has.
\item The fusion of two symmetric phases results in a symmetric phase.
\item The fusion of two symmetry breaking phases results in a symmetry breaking phase but with a nontrivial coefficient.
\end{enumerate}
These fusion results hold even when the symmetry is only partially broken. We explain each case in the following subsections.

\begin{figure}[ht]
    \centering
    \includegraphics[scale=0.5]{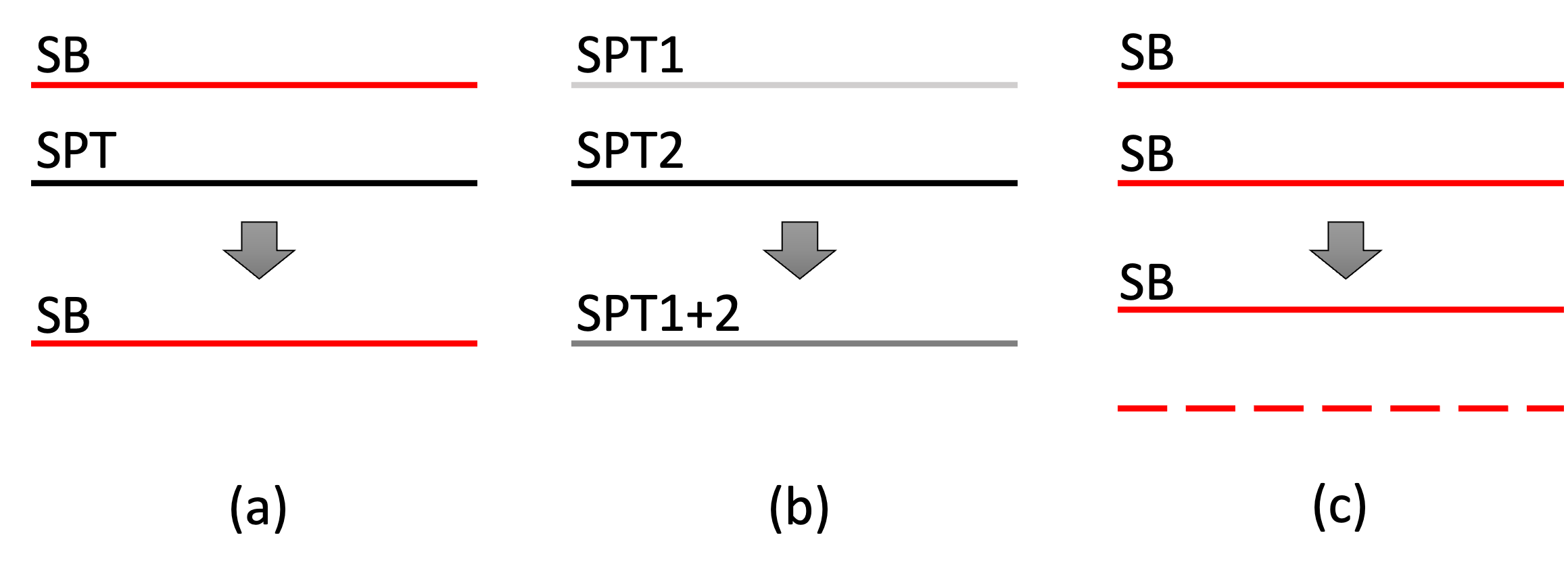}
    \caption{Fusion result of 1D gapped phases. (a) SB and SPT fuse into SB. (b) SPT and SPT fuse into SPT. (c) SB and SB fuse into SB with nontrivial coefficient (the dashed line).}
    \label{fig:fusion}
\end{figure}

\subsection{SB $\times$ SPT}
\label{sec:SBxSPT}

To demonstrate the fusion circuit in this case, we are going to use the symmetry-breaking phase on the upper chain and the symmetric phase of the 1D Ising model on the lower chain with the Hamiltonian
\begin{equation}
H^u = -\sum_i Z^u_iZ^u_{i+1} \text{\ and\ } H^l = -\sum_i X^l_i 
\end{equation} 
where the superscripts $l$ and $u$ denote operators acting on the upper and lower chain.
The two chains can be fused with the circuit shown in Fig.~\ref{fig:SB+SY}. The first step involves $R(ZZ)$ gates on all vertical pairs connected by green dashed lines. The second step involves $R(X)$ gates on the qubits in the lower chain. The Hamiltonian terms in the symmetry breaking chain remain invariant while the $-X^l_i$ Hamiltonian terms in the symmetric chain are mapped to $-Z_i^uZ_i^l$ terms between each vertical pair of qubits. 
The wavefunction transforms as 
\begin{equation}
\begin{aligned}
 & \left(|00...0\rangle + |11...1\rangle \right) \otimes |++...+\rangle \\
\to & |00...0\rangle \otimes |00...0\rangle + |11...1\rangle \otimes |11...1\rangle\, .
\end{aligned}
\end{equation}
That is, the tensor product of a SB GHZ state in one chain and a symmetric product state in the other is mapped to a GHZ state on two chains such that the order parameters from the two chains match.

\begin{figure}[ht]
    \centering
    \includegraphics[scale=0.8]{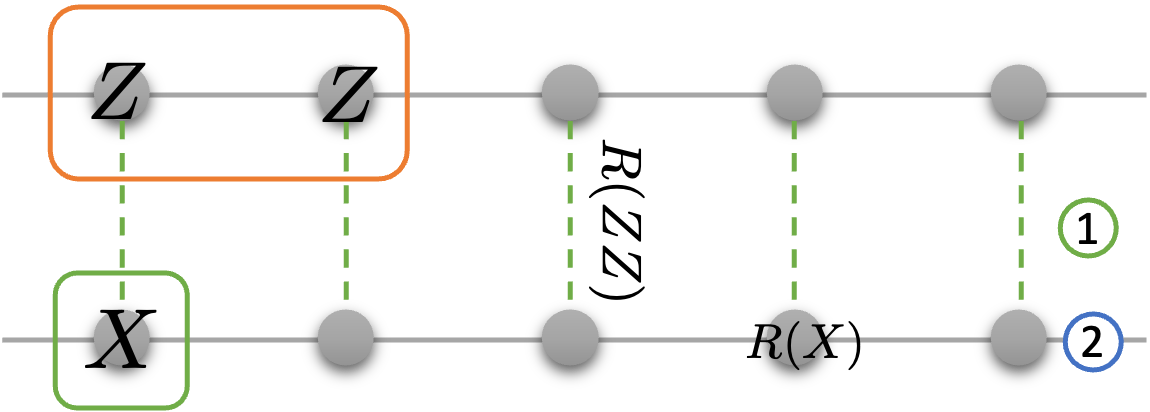}
    \caption{Fusion of the symmetry breaking ground state and the symmetric ground state of the Ising model into the symmetry breaking state with a symmetric finite depth circuit. The first step involves $R(ZZ)$ gates on all vertical pairs connected by green dashed lines. The second step involves $R(X)$ gates on the qubits in the second line.}
    \label{fig:SB+SY}
\end{figure}

Of course, if we think of these gapped phases as defects within a higher dimensional trivial symmetric bulk state, the symmetric phase of the Ising chain is a trivial defect and nothing needs to be done to fuse it with the symmetry breaking defect. Instead, we can view the inverse of the above circuit as a way to map the GHZ state on two chains to the GHZ state on one chain. This step can be combined with the circuits discussed in the following subsections to put the 2-chain symmetry breaking state into the standard form on one chain only. 

When the symmetric phase has a nontrivial SPT order, it corresponds to a nontrivial defect in the higher dimensional trivial symmetric bulk. Its fusion with a symmetry breaking phase still results in a symmetry breaking state, as demonstrated below with the symmetry breaking phase 
\begin{equation}
H^u = \sum_{i} -Z^u_iZ^u_{i+1}-Z^u_{i-\frac12}Z^u_{i+\frac12}
\end{equation}
and the cluster state 
\begin{equation}
H^l = \sum_i - Z^l_{i-\frac12}X^l_iZ^l_{i+\frac12} - Z^l_iX^l_{i+\frac12}Z^l_{i+1}
\end{equation}
with $Z_2\times Z_2$ symmetry. The circuit, as shown in Fig.~\ref{fig:SB+SPT}, keeps the Hamiltonian terms in the symmetry breaking chain invariant while mapping the Hamiltonian terms in the cluster state chain to vertical $-Z^uZ^l$ pairs, hence resulting in a combined symmetry breaking state of $Z_2\times Z_2$ on two chains. Applying (two copies of) the inverse of the circuit in Fig.~\ref{fig:SB+SY}, we can further map it back to the $Z_2\times Z_2$ symmetry breaking state on one chain with symmetric product state in the other.

\begin{figure}[ht]
    \centering
    \includegraphics[scale=0.7]{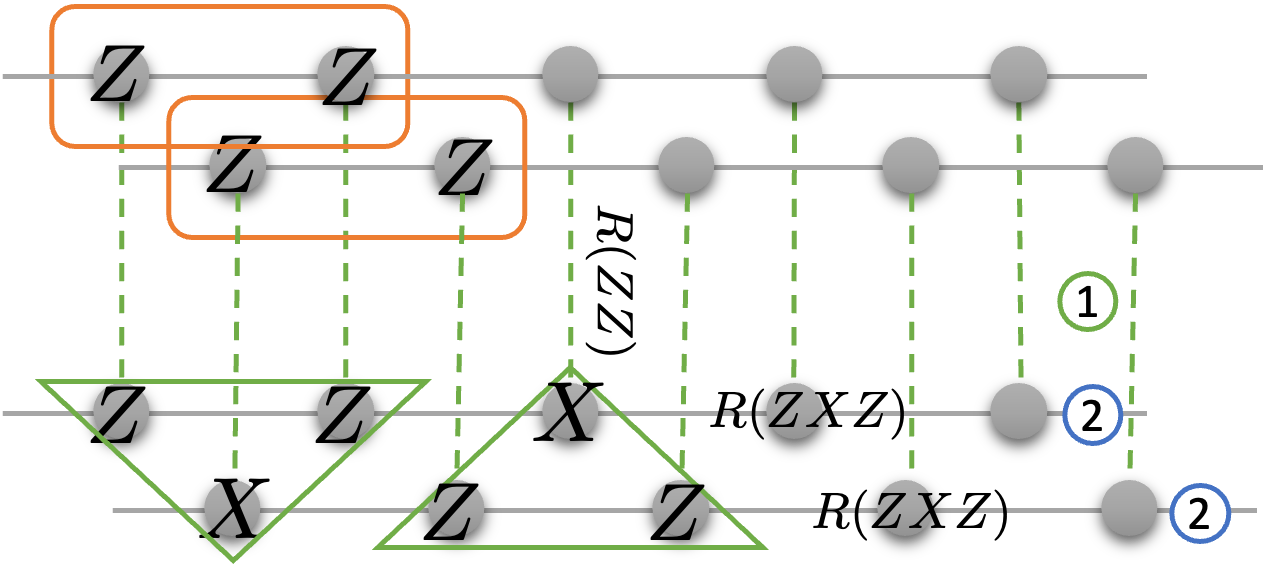}
    \caption{Fusion of the symmetry breaking ground state and the nontrivial SPT state with $Z_2\times Z_2$ symmetry into the symmetry breaking state with a symmetric finite depth circuit. The first step involves $R(ZZ)$ gates on all vertical pairs connected by green dashed lines. The second step involves $R(ZXZ)$ gates generated by the $ZXZ$ Hamiltonian terms in the lower chains (green triangles).}
    \label{fig:SB+SPT}
\end{figure}

To summarize, we find
\begin{equation}
\text{SB} \times \text{SPT} \to SB\, .
\end{equation}

\subsection{SPT $\times$ SPT}
\label{sec:SPTxSPT}

It is well known that SPT phases under stacking form an abelian group given by $H^2(G,U(1))$ \cite{Chen2011classification,Schuch2011}. We briefly review the argument. 

\begin{figure}[ht]
    \centering
    \includegraphics[scale=0.5]{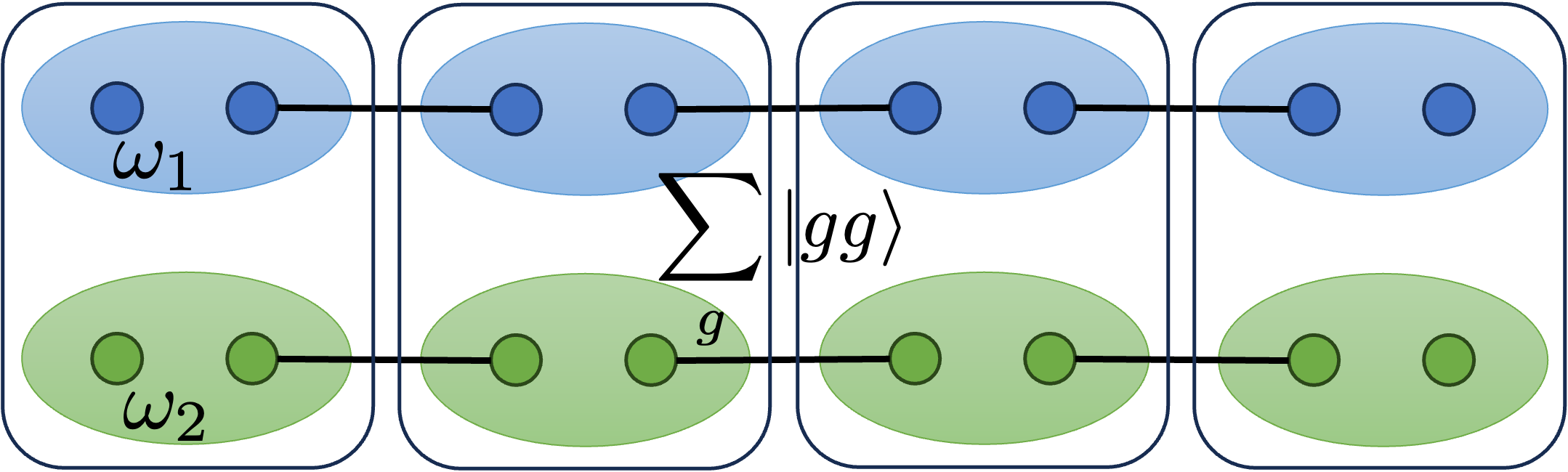}
    \caption{Fusion of two SPT states into one SPT state. The two SPT states have projective edge states given by $\omega_1$ and $\omega_2$ respectively. The fusion of the two SPTs gives a third SPT with edge state given by $\omega_1\omega_2$.}
    \label{fig:SPT+SPT}
\end{figure}

Take two SPT phases in the standard form of two projective representations on site. The left spin with basis states $|g\rangle, g\in G$, transforms under the left half of the symmetry as 
\begin{equation}
V_L(g')|g\rangle = \omega(g',g)|g'g\rangle
\end{equation}
while the right spin transforms under the right half of the symmetry as
\begin{equation}
V_R(g')|g\rangle = \omega^*(g',g)|g'g\rangle\, .
\end{equation}
The onsite symmetry $U(g)= V_L(g)\otimes V_R(g)$ forms a linear representation of $G$, 
\begin{equation}
U(g_1)U(g_2) = U(g_1g_2)
\end{equation}
while $V_L$ and $V_R$ each forms a projective representation
\begin{equation}
\begin{aligned}
V_L(g_1)V_L(g_2) &= \omega(g_1,g_2)V_L(g_1g_2)\\
V_R(g_1)V_R(g_2) &= \omega^*(g_1,g_2)V_R(g_1g_2)\, .
\end{aligned}
\end{equation}
The pair of spins connected between nearest neighbor sites are in the entangled state of
\begin{equation}
\sum_g |gg\rangle\, .
\end{equation}
To map the two chains to one SPT chain, we can simply map the two entangled pairs in the upper and lower chain into a `diagonal' entangled state
\begin{equation}
\left(\sum_{g^u} |g^ug^u\rangle\right) \otimes \left(\sum_{g^l} |g^lg^l\rangle\right) \to \sum_{\tilde{g}} |\tilde{g}\tilde{g}\rangle
\end{equation}
where $\tilde{g} = g^ug^l$. Each step like this is symmetric. Therefore, we find a finite depth symmetric circuit to map the two SPT chains into one with projective edge state associated with
\begin{equation}
\tilde{w}(g_1,g_2) = \omega^u(g_1,g_2)\omega^l(g_1,g_2)\, .
\end{equation}
This can be seen from 
\begin{equation}
\tilde{V}(g')|\tilde{g}\rangle = V^u(g')V^l(g')|g^ug^l\rangle = \omega^u(g',g)\omega^l(g',g)|\tilde{g'g}\rangle
\end{equation}
and 
\begin{equation}
\tilde{V}(g_1)\tilde{V}(g_2) = \omega^u(g_1,g_2)\omega^l(g_1,g_2)\tilde{V}(g_1g_2)\, .
\end{equation}
Therefore, we can write 
\begin{equation}
\text{SPT}_1 \times \text{SPT}_2 \to \text{SPT}_{1+2}
\end{equation}
where the sum in the subscript $1+2$ corresponds to the abelian composition of projective representations in $H^2(G,U(1))$.

\subsection{SB $\times$ SB}
\label{sec:SB+SB}

The fusion of two symmetry breaking phases results in a symmetry-breaking phase with a nontrivial fusion coefficient. Let's discuss this case carefully.

Consider two chains both in the symmetry breaking phase of the Ising chain
\begin{equation}
H^u = -\sum_i Z^u_iZ^u_{i+1}, \ \ H^l = -\sum_i Z^l_iZ^l_{i+1}\, .
\end{equation}
We take the wave-function in both chains to be the symmetrized GHZ state while keeping in mind that there is another degenerate state with nontrivial total charge.

\begin{figure}[ht]
    \centering
    \includegraphics[scale=0.7]{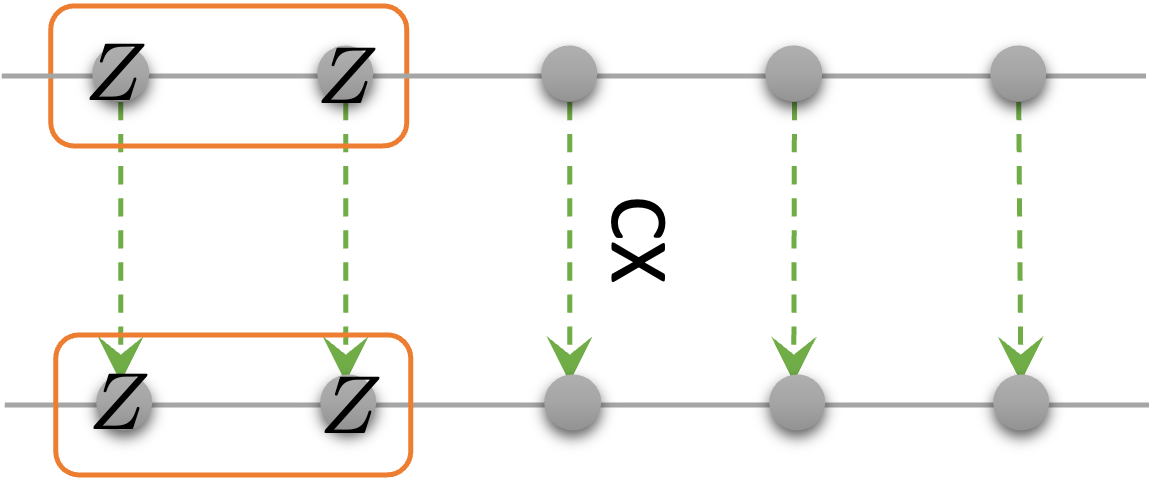}
    \caption{Fusion of two symmetry breaking phases. The circuit consists of pairwise contolled-NOT (CX) gates with the spin in the first chain as control and the corresponding spin the second chain as target.}
    \label{fig:SB+SB}
\end{figure}

The circuit shown in Fig.~\ref{fig:SB+SB} fuses the two chains together. The circuit is composed of controlled-Not gates between pairs of spins in the two chains, defined as $CX^{ct} = |0\rangle\langle 0|_c \otimes I_t + |1\rangle\langle 1|_c\otimes X_t$ acting on control (c) and target (t) qubits. The spins in the top chain are used as control while the spins in the bottom chain are used as target. The controlled-Not gates are not symmetric. Instead, we are going to think of it as implementing a local change of basis and let the symmetry operator transform with it.
\begin{equation}
\prod_i CX^{ul}_i \prod_i X^u_iX^l_i \prod_i CX^{ul}_i = \prod_i X^u_i\, .
\end{equation}
The wave function remains invariant but now the interpretation is different. 
\begin{equation}
\begin{aligned}
 & \left(|00...0\rangle + |11...1\rangle \right) \otimes \left(|00...0\rangle + |11...1\rangle\right) \\
\to & \left(|00...0\rangle + |11...1\rangle \right) \otimes \left(|00...0\rangle + |11...1\rangle \right)\, .
\end{aligned}
\end{equation}
The transformed symmetry operator acts on the first chain only and the first GHZ state is the fusion result, which is still in the SB phase. The second chain does not transform under the symmetry any more. It is also in the GHZ state, but with a degeneracy not protected by the symmetry. This is the coefficient in front of the fusion result
\begin{equation}
\text{SB} \times \text{SB} \to \mathcal{Z}_2 \times \text{SB}
\end{equation}
and is represented by the dashed red line in Fig.~\ref{fig:fusion} (c). This coefficient is going to play an interesting role in the fusion of symmetry breaking phases with domain walls, as we discuss in section~\ref{sec:fusion+}.

\section{Fusion of domain walls}
\label{sec:fusion_dw}

Domain walls on the same 1D chain can be fused with 0D symmetric local unitary gates when they are at a finite distance from each other. The result is summarized in Fig.~\ref{fig:fusion_dw}. 

\begin{figure}[ht]
    \centering
    \includegraphics[scale=0.6]{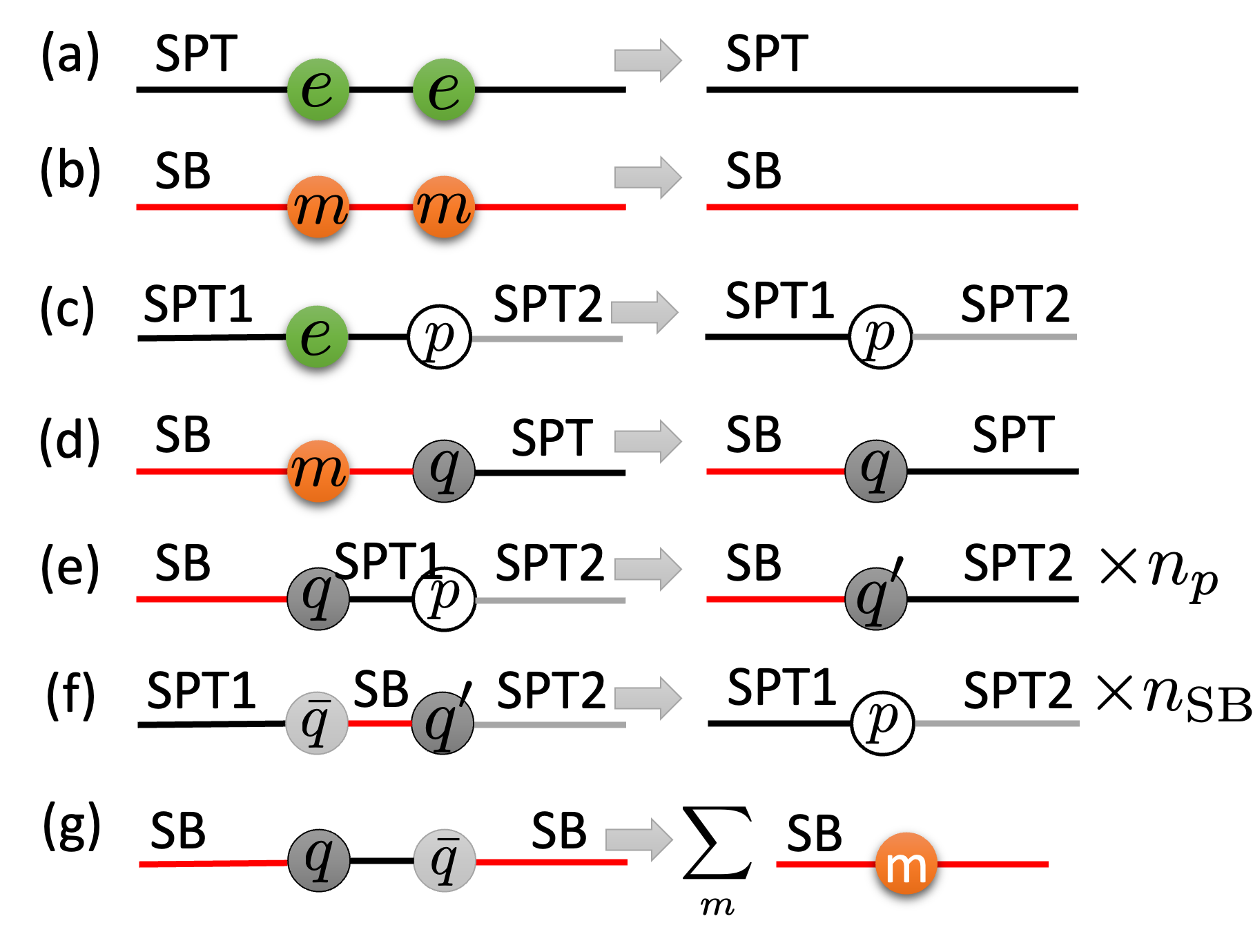}
    \caption{Fusion of domain walls. (a) Fusion of charge domain walls. (b) Fusion of flux domain walls. (c) Fusion of a charge with a projective domain wall into the same projective domain wall between SPT phases. (d) Fusion of a flux domain wall with a $q$ domain wall from SB to SPT phase into the same $q$ domain wall. (e) A $q$ and $p$ domain wall fuse into a $q'$ domain wall with a degeneracy given by the $p$ domain wall. (f) A $\bar{q}$ and $q'$ domain wall fuse into a $p$ domain wall with a degeneracy given by the SB phase. (g) A $q$ and $\bar{q}$ domain wall fuse into all possible flux domain walls.}
    \label{fig:fusion_dw}
\end{figure}

Some of these fusion results are straightforward to see. For example, case (a) and (b)
\begin{equation}
\begin{aligned}
&\text{SPT}-e-\text{SPT}-e-\text{SPT} \to \text{SPT} \\
&\text{SB}-m-\text{SB}-m-\text{SB} \to \text{SB}\, .
\end{aligned}
\end{equation}
We are only going to discuss cases (c) through (f) in detail below.

\subsection{SPT1-e-SPT1-p-SPT2}

Let's consider case (c) where a symmetry charge is fused into the projective edge state between different SPT phases.

Suppose the right half of the system is in the trivial symmetric phase of $Z_2\times Z_2$ symmetry while the left half of the system is in the nontrivial SPT phase,
\begin{equation}
\begin{aligned}
H = &\sum_{i>0} - X_i - X_{i-\frac12} \\
 +&\sum_{i<0} - Z_{i-\frac12}X_iZ_{i+\frac12} - Z_iX_{i+\frac12}Z_{i+1}\, .
\end{aligned}
\label{eq:SYrH2}
\end{equation}
The domain wall is acted upon by a pair of anti-commuting operators
\begin{equation}
Z_{-\frac12}X_0, \ Z_0
\label{eq:pH1}
\end{equation}
neither of which is symmetric.

With a symmetry charge on the trivial symmetric side, the Hamiltonian term at, for example $i=1$ changes sign
\begin{equation}
\begin{aligned}
H =\  &2X_1 + \sum_{i>0} - X_i - X_{i-\frac12} \\
+ &\sum_{i<0} - Z_{i-\frac12}X_iZ_{i+\frac12} - Z_iX_{i+\frac12}Z_{i+1}\, .
\end{aligned}
\label{eq:SYerH2}
\end{equation}
To map Eq.~\ref{eq:SYerH2} back to Eq.~\ref{eq:SYrH2} without the changing degenerate space described by Eq.~\ref{eq:pH1}, we can use the local symmetric unitary operator
\begin{equation}
Z_0Z_1\, .
\end{equation}
In this way, we fuse a $p$ domain wall between two SPT phases with an $e$ domain wall on one of the SPT phase into a $p$ domain wall.
\begin{equation}
\text{SPT}_1-e-\text{SPT}_1-p-\text{SPT}_2 \to \text{SPT}_1-p-\text{SPT}_2\, .
\end{equation}

\subsection{SB-m-SB-q-SPT}

Let's discuss case (d) where an $m$ domain wall on a symmetry-breaking state fuses into the $q$ domain wall between a symmetry-breaking and SPT state. The fusion result has to be a $q$ domain wall. Let's see how that happens through local unitary transformations on the domain walls. 

Consider the $Z_2\times Z_2$ symmetry-breaking state on the right half and the cluster state on the left half
\begin{equation}
\begin{aligned}
H = &\sum_{i>0} -Z_iZ_{i+1} - Z_{i-\frac12}Z_{i+\frac12} \\
 +&\sum_{i<0} - Z_{i-\frac12}X_iZ_{i+\frac12} - Z_iX_{i+\frac12}Z_{i+1}
\end{aligned}
\label{eq:SBrH2}
\end{equation}
and a coupling term
\begin{equation}
Z_{-\frac12}X_0Z_{\frac12}\, .
\label{eq:qH2}
\end{equation}
A domain wall on the SB side corresponds to flipping the sign of one of the $ZZ$ terms, say between $\frac12$ and $3/2$. 
\begin{equation}
\begin{aligned}
H =\  &2Z_{\frac12}Z_{2/3}+ \sum_{i>0} -Z_iZ_{i+1} - Z_{i-\frac12}Z_{i+\frac12}\ \\
& +\sum_{i<0} - Z_{i-\frac12}X_iZ_{i+\frac12} - Z_iX_{i+\frac12}Z_{i+1}\, .
\end{aligned}
\label{eq:SBmrH2}
\end{equation}
To map Eq.~\ref{eq:SBmrH2} back to Eq.~\ref{eq:SBrH2} without changing Eq.~\ref{eq:qH2}, we can use the local symmetric unitary operator
\begin{equation}
Z_0X_{\frac12}Z_1\, .
\end{equation}
In this way, we fuse a $q$ domain wall between the SB and SPT phase with an $m$ domain wall on the SB phase into a $q$ domain wall.
\begin{equation}
\text{SB}-m-\text{SB}-q-\text{SPT} \to \text{SB}-q-\text{SPT}\, .
\end{equation}

\subsection{SB-q-SPT1-p-SPT2}

Consider case (e) where the $q$ domain wall from an SB phase to an SPT phase is fused with the $p$ domain wall from the first SPT phase to another SPT phase. We expect the two to fuse into a single $q$ domain wall but with the multiplicity of the $p$ domain wall contributing a prefactor to the fusion result.

Suppose that the system has a symmetry breaking state on the left, and cluster state on the right, and a trivial state in the middle
\begin{equation}
\begin{aligned}
H &= \hspace{1.9mm} \sum_{i<-1}-Z_{i-1}Z_i -Z_{i-\frac12}Z_{i+\frac12} \\
&+\sum_{i=-1,0,1} -X_i-X_{i+\frac12}  \\
&+\hspace{3.1mm}  \sum_{i>2} -Z_{i-\frac12}X_iZ_{i+\frac12} - Z_{i-1}X_{i-\frac12}Z_{i+\frac12}\, .
\end{aligned}
\end{equation}
The projective edge state between the two SPT states is acted upon by $X_{2}Z_{5/2}$ and $Z_2$. 
Applying the local symmetric unitary transformation 
\begin{equation}
\prod_{i=-1,0,1}R(X_i)R(Z_{i-1}Z_i)R(X_{i+\frac12})R(Z_{i-\frac12}Z_{i+\frac12})
\end{equation}
maps the Hamiltonian to
\begin{equation}
\begin{aligned}
H = & \sum_{i<2}-Z_{i-1}Z_i -Z_{i-\frac12}Z_{i+\frac12} \\
+& \sum_{i>2} -Z_{i-\frac12}X_iZ_{i+\frac12} - Z_{i-1}X_{i-\frac12}Z_{i+\frac12}\, .
\end{aligned}
\end{equation}
The two fold degeneracy can be separated by the eigenvalue of $Z_1Z_2 = \pm 1$. The two cases differ by the action of $Z_{3/2}X_2Z_{5/2}$. In either case, the $q$ domain wall from symmetry breaking to trivial symmetric phases fuses with the $p$ domain wall between the trivial and nontrivial SPT phases and becomes a $q$ domain wall from the symmetry breaking phase to the nontrivial SPT phases. Therefore,
\begin{equation}
\text{SB}-q-\text{SPT}_1-p-\text{SPT}_2 \to n_p\times \text{SB}-q-\text{SPT}_2
\end{equation}
where $n_p$ is the degeneracy of the projective domain wall between the two SPT phases.

\subsection{SPT1-$\bar{q}$-SB-$q'$-SPT2}

Consider case (f) where a $\bar{q}$ domain wall from one SPT phase to the SB phase is fused with a $q'$ domain wall from the SB phase to another SPT phase. $\bar{q}$ and $q'$ fuse into the $p$ domain wall between SPT1 and SPT2, but with a 2 fold degeneracy that comes from squeezing the intermediate SB region. 

Suppose that the Hamiltonian takes the form
\begin{equation}
\begin{aligned}
H =  &\sum_{i<-1} -X_i-X_{i+\frac12} \\
-& Z_{-1}Z_0 - Z_0Z_1 - Z_{-\frac12}Z_{\frac12} \\
+& \sum_{i>1} -Z_{i-1}X_{i-\frac12}Z_i - Z_{i-\frac12}X_iZ_{i+\frac12}\, .
\end{aligned}
\end{equation}
Applying the local symmetric transformation,
\begin{equation}
R(X_{-1})R(Z_{-1}Z_0)R(X_0)R(Z_0Z_1)R(X_{-\frac12})R(Z_{-\frac12}Z_{\frac12})
\end{equation}
maps the Hamiltonian to
\begin{equation}
\begin{aligned}
H = & \sum_{i<0} -X_i-X_{i+\frac12} -X_0 \\
 +&  \sum_{i>1} -Z_{i-1}X_{i-\frac12}Z_i - Z_{i-\frac12}X_iZ_{i+\frac12}\, .
\end{aligned}
\end{equation}
The low energy space is a direct sum of two parts, one with $X_{1/2}=1$, one with $X_{1/2}=-1$. In the first case, the $\bar{q}$ and $q'$ domain walls fuse into a $p$ domain wall between the two SPT phases. In the second case, there is an extra charge at site $1/2$ which can be merged into the $p$ domain wall with an operator $Z_{1/2}X_1Z_{3/2}$. That is,
\begin{equation}
\text{SPT}_1-\bar{q}-\text{SB}-q'-\text{SPT}_2 \to n_{\text{SB}} \times \text{SPT}_1-p-\text{SPT}_2
\end{equation}
where $n_{\text{SB}}$ is the degeneracy of the symmetry breaking phase.

\subsection{SB-$q$-SPT-$\bar{q}$-SB}

Consider case (g) where a $q$ domain wall from SB to SPT phase is fused with a $\bar{q}$ domain wall from the SPT phase back to the SB phase. After the fusion, the SB phases on the two sides are connected, but there are two possibilities at the domain wall. There is either no nontrivial domain wall in between or there is a flux domain wall in between. The fusion result is the direct sum of these two.

To derive this result, we take the $Z_2$ symmetry breaking phase with the on the two sides with the symmetric phase in between. The Hamiltonian for the whole chain is
\begin{equation}
H = \sum_{i<-1} -Z_{i-1}Z_{i} + \sum_{i=-1,0,1} - X_i + \sum_{i>1} Z_iZ_{i+1}
\end{equation}
with symmetrized wave function
\begin{equation}
\left(|...00\rangle + |...11\rangle\right)\otimes |+++\rangle \otimes \left(|00...\rangle + |11...\rangle\right)\, .
\end{equation}
Applying local unitary gates
\begin{equation}
R(Z_{-2}Z_{-1})R(X_{-1})R(Z_{-1}Z_{0})R(X_{0})R(Z_{0}Z_{1})R(X_{1})
\end{equation}
maps the Hamiltonian to
\begin{equation}
H = \sum_{i\leq 1} -Z_{i-1}Z_{i} + \sum_{i>1} Z_iZ_{i+1}
\end{equation}
and the wave-function to
\begin{equation}
\left(|...00\rangle + |...11\rangle\right)\otimes \left(|00...\rangle + |11...\rangle\right)\, .
\end{equation}
The fusion result is hence a direct sum of two possibilities, one corresponding to $Z_1Z_2=1$, the other corresponding to $Z_1Z_2=-1$ -- an $m$ domain wall. In terms of wavefunctions, the two parts in direct sum are
\begin{equation}
|...0000...\rangle + |...1111...\rangle
\end{equation}
and
\begin{equation}
|...0011...\rangle + |...1100...\rangle\, .
\end{equation}
Therefore,
\begin{equation} \label{eq:sbqsptqsb}
\text{SB}-q-\text{SPT}-\bar{q}-\text{SB} \to \text{SB} + \text{SB}-m-\text{SB}\, .
\end{equation}

\section{Fusion of phases with domain walls}
\label{sec:fusion+}

When the 1D phases to be fused have domain walls on them, they can again be fused with symmetric finite-depth circuits, but we need to be careful in choosing the finite-depth circuit. This is because in many situations domain walls can be created / annihilated with finite depth circuits. For example, the flux domain wall in a symmetry breaking state can be created by applying symmetry to a segment on the chain, which is a depth one circuit. The charge domain wall on symmetric phases can also be created with a finite depth circuit simply by applying the charge hopping operator to a segment. Therefore, in order to properly discuss the fusion of 1D phases with domain wall, we need to use a circuit that preserves the existence of a domain wall. To that end, we can use the circuit discussed in section~\ref{sec:fusion} to fuse the bulk of the gapped phase and add extra symmetric local unitary to fuse the domain walls. That is, on the two sides of the domain walls, we use the same circuit used for fusing gapped phases without domain walls. At the domain wall, we have the freedom to change the circuit by a symmetric local unitary. By doing so, we make sure we do not have the freedom to create / remove domain walls with the fusion circuit. 

Fig.~\ref{fig:fusion+} summarizes the fusion result of gapped phases with domain walls. The results in (a) and (b) are straightforward. We will discuss (c), (d), (e) in the follow subsections. We remark that an interesting subtlety can occur in case (b) when the symmetry is only partially broken, in which the SPT is absorbed at the cost of creating a charge domain wall. This is discussed in Appendix \ref{sec:partialssb}.

\begin{figure}[ht]
    \centering
    \includegraphics[width=\linewidth]{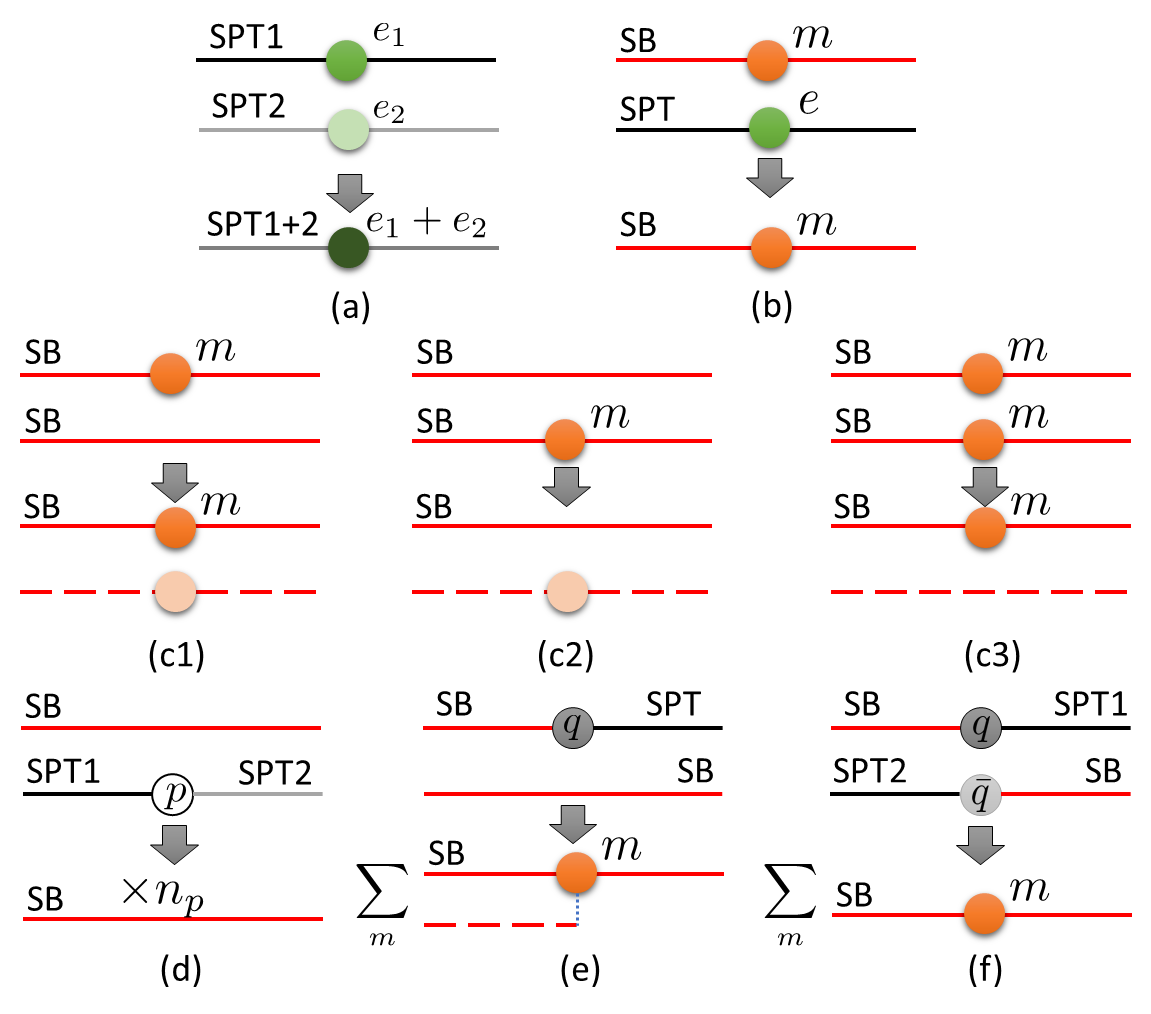}
    \caption{Fusion of 1D gapped phases with domain walls. (a) SPT phases and their charge domain walls fuse according to their additive group structure. (b) Fusion of SB phase with a flux domain wall and SPT phase with a charge domain wall results in SB phase with a flux domain wall. (c) Fusion of two SB phases with or without domain walls; the dashed line is the fusion coefficient. (d) Fusion of SB phase with two segments of SPT phases with a projective domain wall in between; $n_p$ is the degeneracy of the projective domain wall. (e) Fusion of a chain half in SB phase and half in SPT phase with a SB chain results in a SB chain with all possible flux domain walls summed over. The dotted line between the domain wall and the fusion coefficient indicate their coupling. (f) Fusion of two chains with $q$ and $\bar{q}$ domain walls.}
    \label{fig:fusion+}
\end{figure}

\subsection{SB $\times$ SB}
\label{sec:SBxSB_+}

Let's discuss case (c) of a symmetry breaking phase fused with another symmetry breaking phase carefully when one or both of them carry flux domain walls. As shown in Fig.~\ref{fig:fusion+} (c), the fusion result might depend on the ordering of the defects to be fused. 

First, let's consider the case where the first chain has a flux defect while the second does not. The wave function before fusion is
\begin{equation}
\left( |...0011...\rangle + |...1100...\rangle \right) \otimes \left( |...0000...\rangle + |...1111...\rangle \right)\, .
\end{equation}
Applying the same fusion circuit as in Section~\ref{sec:SB+SB} when two symmetry breaking phases without domain walls are fused, we get
\begin{equation}
\left( |...0011...\rangle + |...1100...\rangle \right) \otimes \left( |...0011...\rangle + |...1100...\rangle \right)\, .
\end{equation}
As the transformed symmetry operator only acts on the first chain, we interpret the first chain as the fusion result while the second chain is the coefficient. We see that the fusion result is a symmetry breaking chain with flux domain wall, while the coefficient is also in a GHZ state with domain wall. 

When the two chains are exchanged, the fusion has a different result. 
\begin{equation}
\left( |...0000...\rangle + |...1111...\rangle \right) \otimes \left( |...0011...\rangle + |...1100...\rangle \right)
\end{equation}
is mapped to
\begin{equation}
\left( |...0000...\rangle + |...1111...\rangle \right) \otimes \left( |...0011...\rangle + |...1100...\rangle \right)\, .
\end{equation}
The fusion result is hence the symmetry breaking chain without domain wall, while the coefficient stays the same as the previous case which is a GHZ state with domain wall.

Finally, when both chains contain flux domain walls,
\begin{equation}
\left( |...0011...\rangle + |...1100...\rangle \right) \otimes \left( |...0011...\rangle + |...1100...\rangle \right)
\end{equation}
it maps to
\begin{equation}
\left( |...0011...\rangle + |...1100...\rangle \right) \otimes \left( |...0000...\rangle + |...1111...\rangle \right)\, .
\end{equation}
The fusion result is a symmetry breaking chain with domain wall, while the coefficient does not have a domain wall. 

To summarize, we find
\begin{equation}
\begin{aligned}
&\text{SB}-m-\text{SB} \times \text{SB} \to \mathcal{Z}_2-m-\mathcal{Z}_2 \times \text{SB}-m-\text{SB} \\
&\text{SB} \times \text{SB}-m-\text{SB} \to \mathcal{Z}_2-m-\mathcal{Z}_2 \times \text{SB} \\
&\text{SB}-m-\text{SB} \times \text{SB}-m-\text{SB} \to \mathcal{Z}_2 \times \text{SB}-m-\text{SB}\, .
\end{aligned}
\end{equation}

\subsection{SB $\times$ SPT-p-SPT}

Consider case (d) involving the fusion of a symmetry breaking chain with an SPT chain but with two different SPT orders on the two sides and a $p$ domain wall in between. As the SB order `eats up' the SPT orders, we expect the $p$ domain wall to disappear after the fusion. 

\begin{figure}[ht]
    \centering
    \includegraphics[scale=0.65]{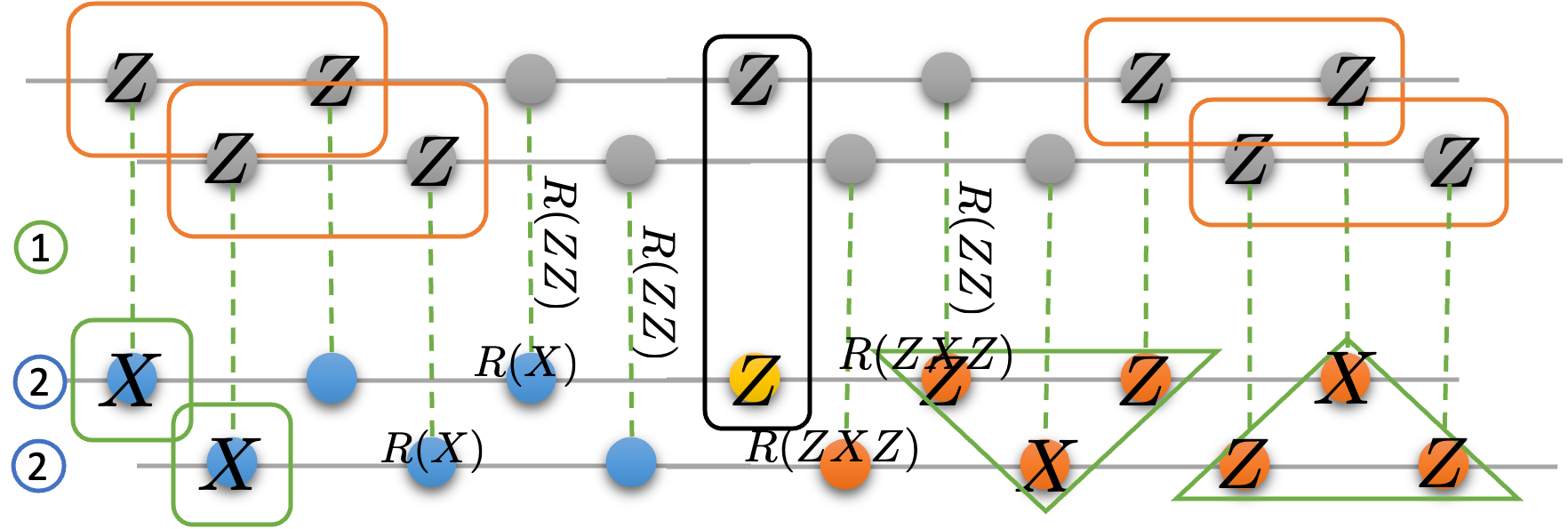}
    \caption{Fusion of SB and SPT phases in the presence of a $p$ domain wall.}
    \label{fig:SB+SPT-p-SPT}
\end{figure}

To see how explicitly that happens, consider the setup in Fig.~\ref{fig:SB+SPT-p-SPT} with a top chain with $Z_2\times Z_2$ symmetry breaking order. The Hamiltonian in the top chain is given by
\begin{equation}
H^u = \sum_i -Z^u_iZ^u_{i+1}-Z^u_{i-\frac12}Z^u_{i+\frac12} \, .
\end{equation}
The bottom chain has trivial SPT order on the left hand side and nontrivial SPT order on the right hand side. The Hamiltonian in the bottom chain is given by
\begin{equation}
\begin{aligned}
H^l =  & \sum_{i<0} -X_i^l - X_{i+1/2}^l \\
  +&\sum_{i>0} -Z^l_{i-\frac12}X^l_iZ^l_{i+\frac12}-Z^l_iX^l_{i-\frac12}Z^l_{i+1}\, .
\end{aligned}
\end{equation}
On the two sides of the domain wall, we can use the circuit in Fig.~\ref{fig:SB+SY} and Fig.~\ref{fig:SB+SPT} respectively to fuse the trivial and non-trivial SPTs into the symmetry breaking state. In particular, step 1 involves the gate $R(ZZ)$ on each vertical pair connected by dashed green lines, and step 2 involves $R(X)$ on the blue dots and $R(ZXZ)$ centered on the orange dots in the lower chain. After these two steps, the horizontal $ZZ$ terms remain while the $X$ and $ZXZ$ terms are replaced by vertical $ZZ$ terms on pairs of spins connected by the green dashed line. Therefore, all qubits are merged into the symmetry-breaking state except for the yellow dot. 

The degeneracy associated with the edge state on the domain wall can be removed by the $ZZ$ term in the black box, which commutes with all other terms in the Hamiltonian. The low energy space of the system is a direct sum of two parts, one with eigenvalue $+1$ under the $ZZ$ term in the black box, the other with a $-1$ eigenvalue. With either eigenvalue, the yellow dot merges into the symmetry breaking state and the two differ only by the local symmetric operation of $X$ on the yellow dot. Therefore, there is a factor of 2 in the fusion result in Fig.~\ref{fig:SB+SPT-p-SPT}, but otherwise we see the projective edge state between two SPT states disappears when fused with a symmetry-breaking state. That is,
\begin{equation}
\text{SB} \times \text{SPT}_1-p-\text{SPT}_2 \to n_p \times \text{SB}\, .
\end{equation}

\subsection{SB-$q$-SPT $\times$ SB}

Consider case (e) where a SB-SPT chain is fused with a SB chain. We expect the whole system to fuse into a SB state but with the possibility of having or not having a $m$ domain wall in between. 

Suppose that both chains have a $Z_2$ symmetry. 
The upper chain contains a symmetry breaking phase on the left half and a symmetric phase on the right half,
\begin{equation}
\begin{aligned}
H^u = - \sum_{i<0} Z^u_{i}Z^u_{i+1} - \sum_{i>0} X^u_i\, .
\end{aligned}
\end{equation}
The lower chain is in a symmetry breaking phase,
\begin{equation}
H^l = \sum -Z^l_iZ^l_{i+1}\, .
\end{equation}
Fusion can be realized with a circuit shown in Fig.~\ref{fig:SB-SPT+SB}. 

\begin{figure}[ht]
    \centering
    \includegraphics[scale=0.65]{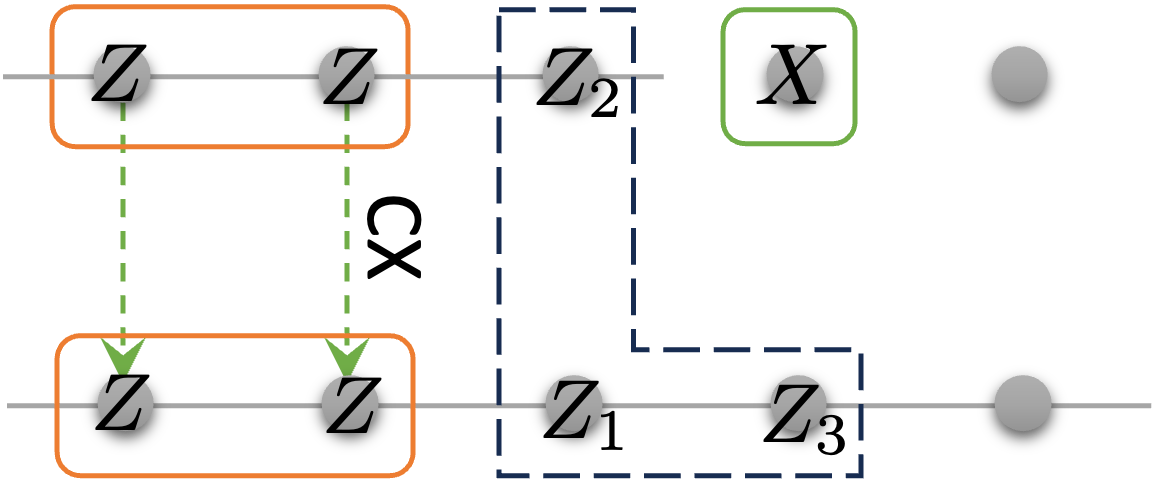}
    \caption{Fusion of a SB-SPT chain with a $q$ domain wall and a SB chain.}
    \label{fig:SB-SPT+SB}
\end{figure}

The orange and green boxes indicate the Hamiltonian terms before the fusion. The circuit on the left-hand side follows the one in Fig.~\ref{fig:SB+SB} and is composed of CX gates from the top chain to the bottom chain. 

After the circuit, all the Hamiltonian terms remain invariant on the two sides of the domain wall. Global $Z_2$ symmetry does not act on the left half of the lower chain any more and it becomes a fusion coefficient. The left half of the upper chain and the right half of the lower chain merge into a SB chain which is the fusion result. At the domain wall, there is now a three body $Z_1Z_2Z_3$ term, indicated by the dashed black box in Fig.~\ref{fig:SB-SPT+SB}. This term couples the order parameter of the fusion coefficient $Z_1$ with the domain wall on the fusion result $Z_2Z_3$. 

Therefore, we can write
\begin{equation}
\text{SB}-q-\text{SPT} \times \text{SB} \to \sum_m \text{SB}-m-\text{SB}\, .
\end{equation}
although this way of writing does not make explicit the coupling of $m$ to the fusion coefficient.

\subsection{Other fusions}

Other more complex fusions can be understood in terms of simpler fusions using the following trick. In Fig.~\ref{fig:fusion_trick}, we show how a fusion of two systems, each containing a domain wall, can be understood as two separate phase fusions, each involving only a single domain wall, followed by a fusion of domain walls. 

\begin{figure}
    \centering
    \includegraphics[width=\linewidth]{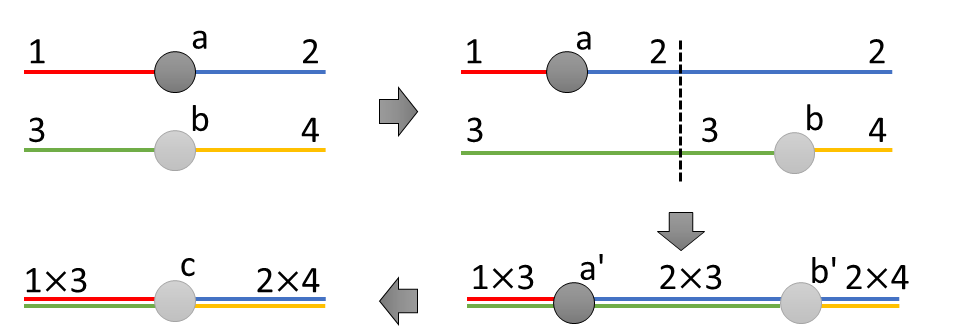}
    \caption{Fusion of phases 1 and 2 separated by a domain wall $a$ and phases 3 and 4 separated by a domain wall $b$. We can first separate the domain walls a bit to the left and right, which doesn't change anything. Then we fuse the two systems on either side of the dotted line independently (each containing only one domain wall). This results in a 1D system with two domain walls $a'$ and $b'$ separated by an intermediate phase ($2\times 3$), which can then be fused into one domain wall $c$. For simplicity, we do not draw fusion coefficients, degeneracy, or sums over domain walls, all of which are possible in general.}
    \label{fig:fusion_trick}
\end{figure}

As an application of this trick, consider the fusion $\text{SB}-q-\text{SPT}_1 \times \text{SPT}_2-\bar{q}-\text{SB}$ shown in Fig.~\ref{fig:fusion+}(f). 
First, we note the simple fusion result
\begin{equation}
    \text{SB}-q-\text{SPT}_1 \times \text{SPT}_2 \to \text{SB}-q-\text{SPT}_{1+2}
\end{equation}
which can be explicitly obtained using similar circuits as in previous examples. Using this, and applying the trick in Fig.~\ref{fig:fusion_trick}, we find
\begin{equation}
\begin{aligned}
\text{SB}-q-\text{SPT}_1 &\times \text{SPT}_2-\bar{q}-\text{SB} \\
&\to \text{SB}-q-\text{SPT}_{1+2}-\bar{q}-\text{SB}\\
&\to \sum_m \text{SB}-m-\text{SB}\, .
\end{aligned}
\end{equation}
where the last step used Eq.~\ref{eq:sbqsptqsb} to fuse the $q$ and $\bar{q}$ domain walls.

\section{Some results in higher dimensions}
\label{sec:2D}

Gapped phases in two or higher dimensions form categories of even higher order \cite{Johnson-Freyd2022,Kong2015arxiv,Kong2022}. For example, in 2D gapped phases there are 1D domain walls and there can further be 0D domain walls on top of the 1D domain walls. A complete discussion of the fusion of higher dimensional gapped phases and their domain walls is much more complicated and beyond the scope of this paper, but we do want to discuss a few simple cases, and show how things work in an analogous way as their 1D counterparts.

\subsection{1D domain walls in 2D SPTs}

Consider the 2D SPT state with $Z_2$ symmetry. One type of 1D defect is a symmetry breaking defect where the symmetry charge condenses. One can then ask if there are any other types of interesting 1D defects. In this section we discuss whether a 1D symmetry defect line is a nontrivial defect inside the SPT state. That is, suppose the $Z_2$ symmetry is applied to a region inside the $2D$ state. As the state is symmetric, it remains invariant both inside and outside the region, but can change along the boundary. In this section, we are going to show using the fixed-point form of the SPT state discussed in Ref.~\onlinecite{Chen2011,Chen2013} that the changes along the boundary can be induced with a finite depth symmetric circuit. Therefore, a symmetry defect line is not a nontrivial 1D defect in the SPT state.

\begin{figure}[ht]
    \centering
    \includegraphics[scale=0.5]{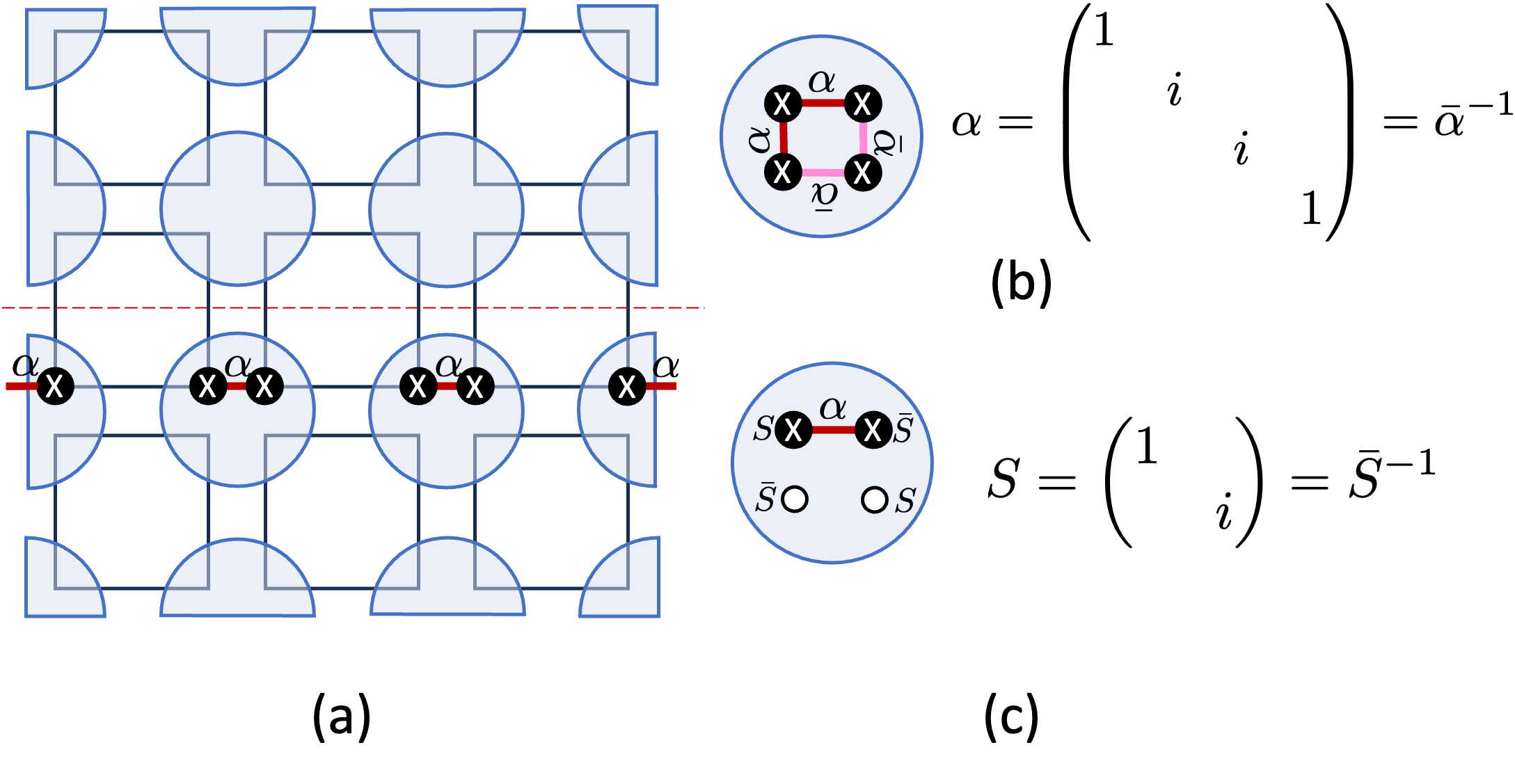}
    \caption{Symmetry defect in 2D $Z_2$ SPT state. Each lattice site (blue discs) hosts four qubits. Every four qubits connected into a square are in the $|0000\rangle+|1111\rangle$ state. The $Z_2$ symmetry action on each site is given in (b). Applying the symmetry in the lower half plane induces transformation along the boundary line (dotted red) as shown in (a). The same transformation can be induced by applying the gate set in (c) to each site which are $Z_2$ symmetric.}
    \label{fig:2DSPT}
\end{figure}

Consider the 2D state as shown in Fig.~\ref{fig:2DSPT}, where the each lattice site (blue discs) hosts four qubits. The $Z_2$ symmetry on each lattice site is given in (b) which involves $\prod X$ on all the qubits as well as phase factor $\alpha\alpha\bar{\alpha}\bar{\alpha}$ over connected pairs. In the ground state, every four qubits connected into a square are in the local entangled state $|0000\rangle + |1111\rangle$. The wave-function remains invariant if the $Z_2$ symmetry is applied to all lattice sites. If the symmetry is applied only to a subregion (the lower half plane for example), the wave function changes along the boundary of the subregion. The change in the wave function corresponds to applying $\prod X$ to all the black dots in (a) and $\prod \alpha$ to all the red bonds in (a). To realize this change with a symmetric finite depth circuit, we can apply the transformation shown in (c) to the lattice sites along the dotted boundary line. It can be explicitly checked that this realizes the same unitary transformation as (a) and the unitary in each lattice site commutes with the $Z_2$ symmetry and is hence symmetric.

\subsection{Fusion of 1-form symmetry breaking phases}

\begin{figure}[ht]
    \centering
    \includegraphics[scale=0.50]{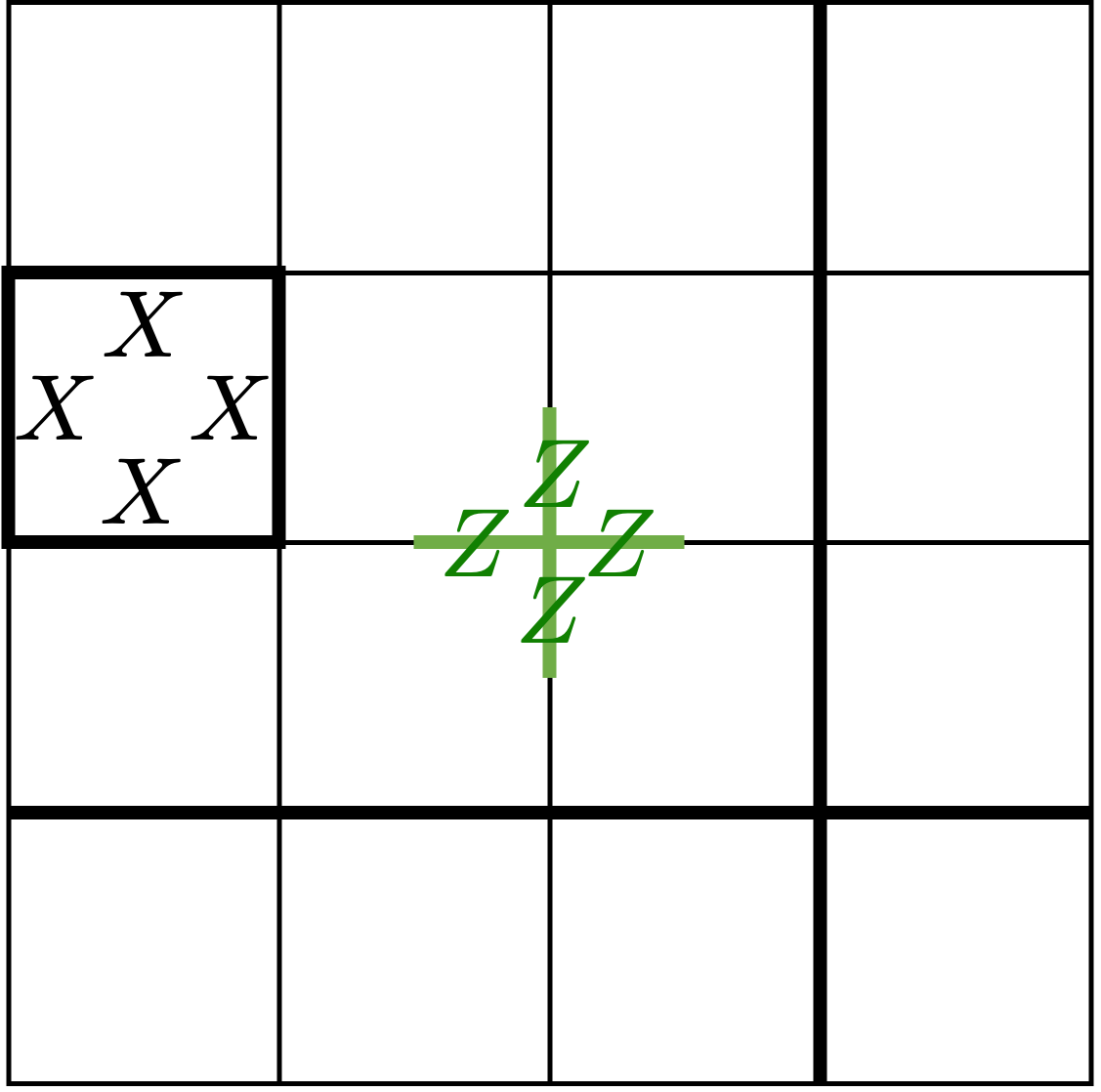}
    \caption{Toric code model on the square lattice. The model has a 1-form symmetry given by $\prod X$ on all closed loops including the nontrivial ones (think horizontal and vertical lines).}
    \label{fig:TC}
\end{figure}

1-form symmetries start to play an interesting role in 2D gapped phases. In particular, the breaking of 1-form symmetry results in topological order. For example, the 2D Toric code with $Z_2$ topological order breaks a $Z_2$ 1-form symmetry of the Wilson lines. The point defect of this symmetry are $Z_2$ gauge charge excitations. We show in this section how the fusion of two 1-form symmetry breaking Toric code states with or without symmetry defect follows similar rules as that discussed in section~\ref{sec:SBxSB_+}  for 1D 0-form symmetry breaking phases with or without flux domain wall. 

Consider the 2D Toric Code state defined on square lattice as shown in Fig.~\ref{fig:TC}. The $Z_2$ gauge field degrees of freedom are on the edges. The Hamiltonian terms include the four-body plaquette terms of $\prod_{e\in p} X_e$ and four-body vertex terms of $\prod_{v\in e} Z_e$. The 1-form symmetry is given by $\prod X$ on all closed loops, including nontrivial loops in the $x$ and $y$ directions. The ground state spontaneously breaks this 1-form symmetry. If the $|0\rangle$ state on each edge is regarded as no string and the $|1\rangle$ state on each edge is regarded as having a string, the symmetrized ground state wave function is an equal weight superposition of all closed-loop configurations, including the non-trivial ones.
\begin{equation}
|\psi\rangle = \sum_{C:\text{closed loop configurations}} |C\rangle
\end{equation}

With two copies of the Toric Code state, there is a $Z_2\times Z_2$ 1-form symmetry. We consider the situation where only the diagonal $Z_2$ 1-form symmetry is preserved, which may come from the 1-form symmetry of a higher dimensional bulk. Applying pair-wise controlled-Not gates between corresponding qubits in the two toric code states keeps the two states invariant but changes the diagonal 1-form symmetry to act on only the first copy of Toric Code. Therefore, after the transformation, two copies of the Toric Code fuse into one copy, with the coefficient also being a Toric Code. 
\begin{equation}
\text{T.C.} \times \text{T.C.} \to \text{t.c.} \times \text{T.C.}
\end{equation}
where the lower case t.c. represents the coefficient state not acted upon by the 1-form symmetry. 

Now consider the situation where either one or both of the Toric Code to be fused has a 1-form symmetry defect. When the action of the 1-form symmetry operator is to add closed loops in the wave function, a 1-form symmetry defect corresponds to end of string, which we label as the $e$ excitation. 

When the first Toric Code has a defect while the second doesn't, applying the controlled-Not circuit adds the end of string to the second wave-function without changing the first wave-function. Therefore, the fusion result has a defect as well as the coefficient (the t.c. state). When the first Toric Code state has no defect while the second one has, the controlled-Not gates do not change either wave-function. Therefore, the fusion result has no defect while the coefficient does. Finally, when both copies of Toric Code has defects, the controlled-Not gates removes the defect in the second copy. Therefore, the fusion result has a defect while the coefficient does not.
This fusion pattern is exactly the same as that in section~\ref{sec:SBxSB_+} when two symmetry breaking states of global symmetry are fused together.

\section{summary}
\label{sec:summary}

In this paper, we explored the higher category structure of 1D gapped phases and their domain walls. We use the rule that for 1D gapped phases, the equivalence relation is given by 1D finite-depth circuits, while for their domain walls, the equivalence relation is given by 0D local unitary operations. We establish fusion relations between 1D gapped phases, their 0D domain walls, as well as 1D gapped phases with domain walls. In particular, we point out that to properly fuse 1D gapped phases with domain walls, we need to use the same circuit as when fusing 1D gapped phases without domain walls except locally around the domain walls. We find that sometimes fusion is not commutative with respect to the two inputs and sometimes the coefficient in front of the fusion result can be a nontrivial 1D gapped system itself. 

Similar equivalence relations can be used to study gapped defects in higher dimensional nontrivial gapped states, including topological states. The fusion rule discussed in this paper has interesting counterparts with defects and their domain walls in gauge theories. On the other hand, 1D gapped phases can show up as the `uninteresting' kind of defect in non-trivial gapped states that do not couple to the bulk. But they can also show up, for example, as fusion coefficients of the `interesting' defects \cite{Roumpedakis2023,Kong2020defects} and hence should be carefully studied for the completeness of the story.

The fusion structure we studied in this paper for 1D gapped phases is of course only part of the story. To characterize the full 2-category structure of 1D gapped phases / defects, we need much more data. Going onto 2D and higher, the gapped phases / defects form categories of even higher dimensions~\cite{Johnson-Freyd2022,Kong2015arxiv,Kong2022}. To explore the general structure of gapped phases / defects, we will need to make use of more powerful classes of unitaries like Sequential Quantum Circuits \cite{Chen2024sequential} and higher dimensional circuits.


\begin{acknowledgments}
We are indebted to inspiring discussions with Wenjie Ji, Liang Kong, Tian Lan, Michael Levin, Shu-Heng Shao, Nathanan Tantivasadakarn, Robijn Vanhove, Xiao-Gang Wen, Jingyu Zhao. DTS and XC are supported by the Simons Collaboration on Ultra-Quantum Matter, which is a grant from the Simons Foundation (651440 DTS; 651438 XC). XC is supported by the Simons Investigator Award (award ID 828078), the Institute for Quantum Information and Matter at Caltech and the Walter Burke Institute for Theoretical Physics at Caltech. 
\end{acknowledgments}
\newpage

\bibliography{references}

\appendix




\section{Matrix product state argument}
\label{appen:MPS}


In this section, we argue the claim of Sec.~\ref{sec:phases} that symmetric phases only have charged domain walls and symmetry breaking phases only have flux domain walls. 

Consider the gapped ground state in either the symmetric or symmetry breaking phases represented as finite-dimensional matrix product state \cite{Perez-Garcia2006},
\begin{equation}
    |\psi\rangle = \sum_{i_1,\dots,i_n} \mathrm{Tr}(A^{i_1}\cdots A^{i_n})|i_1,\dots,i_n\rangle
\end{equation}
A local excitation around site $k$ can be obtained by changing the matrices on a block of sites surrounding $k$ while keeping the matrices on all other sites invariant. After sufficient enlargement of the unit cell, we can assume that only the matrices on site $k$ are changed from $A^i$ to $\tilde{A}^i$,
\begin{equation}
    |\tilde{\psi}\rangle = \sum_{i_1,\dots,i_n} \mathrm{Tr}(A^{i_1}\cdots\tilde{A}^{i_k}\cdots A^{i_n})|i_1,\dots,i_n\rangle
\end{equation}

The ground state in a gapped symmetric phase can be represented with an `injective' set of matrices $A^i$ ($i=1,...,d$) meaning that (after sufficient enlargement of the unit cell) the set of matrices on one site span the whole space of $D\times D$ matrices where $D$ is the dimension of the matrices $A^i$ \cite{Perez-Garcia2006}. Because of this, no matter what the modified matrices $\tilde{A}^i$ are at the domain wall, they can always be represented as linear combinations of the original set of matrices
\begin{equation}
\tilde{A}^i = \sum_{i'} M_{ii'} A^{i'}
\end{equation}
Thus, the domain wall can be generated by acting with the linear operator $M$ on the degree of freedom at site $k$. This operator is not unitary in general, but it can be replaced by a unitary with the same locality (up to exponentially decaying tails). To see this, let $\mathcal{V}$ be the quasi-local unitary which maps the state $|\psi\rangle$ to the fixed-point state of the SPT phase $|\phi\rangle$ that takes the form of a product of entangled pairs of spins \cite{Chen2011classification}. Ignoring tails, we can take $\mathcal{V}$ to be a FDQC. Then, we have,
\begin{equation}
    M_k|\psi\rangle = M_k\mathcal{V}^\dagger |\phi\rangle = \mathcal{V}^\dagger M'_k|\phi\rangle
\end{equation}
where $M'_k = \mathcal{V} M_k \mathcal{V}^\dagger$. Since $\mathcal{V}$ is a FDQC, $M'_k$ will be supported on a finite number of sites near site $k$. Since $|\phi\rangle$ is a tensor product of entangled pairs, we can always find a unitary $U_k$ which has the same support as $M'_k$ and satisfies $M'_k|\phi\rangle = U_k|\phi\rangle$. Applying $\mathcal{V}^\dagger$ to both sites of this equation, we get,
\begin{equation}
    M_k|\psi\rangle = U'_k|\psi\rangle
\end{equation}
where $U'_k=\mathcal{V}^\dagger U_k \mathcal{V}$ is again a local unitary operator supported in a finite interval around site $k$. 

Therefore, every domain wall in an injective MPS can be generated by acting with a local unitary operator. Importantly, depending on the nature of the domain wall, this local unitary operator may not be symmetric, which leaves the possibility of non-trivial domain walls carrying symmetry charge.

For SB phases, the symmetric ground state is represented as a non-injective MPS. Now the matrices $A^i$ are block-diagonal, $A^i=\bigoplus_\alpha A^i_\alpha$ where $\alpha$ label the different short-range entangled ground states that constitute the SB state \cite{Perez-Garcia2006}. The span of the matrices $A^i$ generates the whole space of block-diagonal matrices. Consider the case where the matrices $\tilde{A}^i$ generating the domain wall are also block-diagonal. Then, we can again find a linear operator $M$ acting on site $k$ which creates the domain wall. As before, we can find a FDQC $\mathcal{V}$ mapping $|\psi\rangle$ to the fixed-point of the symmetry breaking phase, which now has the form $\sum_{\alpha} |\phi_\alpha\rangle$ where each $|\phi_\alpha\rangle$ is a product of entangled pairs, and these pairs are supported in different sectors of Hilbert space for different $\alpha$ \cite{Chen2011classification}. Following the above, we write,
\begin{equation}
    M_k|\psi\rangle = \sum_\alpha M_k\mathcal{V}^\dagger|\phi_\alpha\rangle = \sum_\alpha \mathcal{V}^\dagger M'_k|\phi_\alpha \rangle
\end{equation}
Since each $|\phi_\alpha\rangle$ is a product state, we can again find a local unitary $U_k = \bigoplus_\alpha U_{\alpha,k}$ such that $M'_k|\phi_\alpha\rangle = U_{\alpha,k}|\phi_\alpha\rangle$. 
Then, $U'_k=\mathcal{V}^\dagger U_k \mathcal{V}$ is a local unitary that generates the domain wall. 

As before, this local unitary can be charged under the symmetry. In the SB case, however, this charge delocalizes into the SB state such that it is locally undetectable and therefore no longer corresponds to a non-trivial domain wall. Let us show why this is true using MPS. In non-injective MPS, the delocalization of the charge follows from a symmetry of the tensors \cite{Schuch2010}, 
\begin{equation} \label{eq:ginvariant}
    A^i = U_g A^i U_g^\dagger\, .
\end{equation} 
In the abelian case, $U_g$ are unitary matrices of the form $U_g = \bigoplus_\alpha \lambda_\alpha(g) I_{n_\alpha}$ where $\lambda_\alpha(g)$ is a 1D representation of $G$ and $I_{n_\alpha}$ is the $n_\alpha\times n_\alpha$ identity matrix where $n_\alpha$ is the size of block $\alpha$. 
On one hand, the operators $U_g$ are locally undetectable. That is, if we define a domain wall by the matrices $\tilde{A}^i=A^iU_g$, then this change does not affect any local reduced density matrices since Eq.~\ref{eq:ginvariant} can be used to move $U_g$ far away from any local operator acting on the MPS. On the other hand, the operators $U_g$ carry symmetry charge. Since applying the broken symmetry permutes the different short-range entangled states, it must also permute the blocks $\alpha$ in such a way $U_g$ changes by a 1D representation $\lambda_g$ (a charge). This establishes a mapping $\Gamma:g\mapsto \lambda_g$ between $G$ and the group of 1D representations of $G$, which are isomorphic when $G$ is finite abelian. When the symmetry is fully broken, we have $|G|$ blocks, so $U_g$ can be chosen to be semi-regular, meaning it contains every 1D representation of $G$ (with each block transforming under a different representation). Therefore, $\Gamma$ is an isomorphism.

Gathering the above facts, if a domain wall corresponding to the matrices $\tilde{A}^i$ carries a charge $\lambda$, we can modify it as $\tilde{A}^i\mapsto \tilde{A}^iU_g$ for $g=\Gamma^{-1}(\lambda)$ to turn the corresponding linear operator $M$ and its equivalent unitary $U$ into symmetric operators. At the same time, this modification does not change the local reduced density matrix around the domain wall. Therefore, all domain walls created by local unitaries in the SB phase are trivial.

Finally, domains walls defined by matrices $\tilde{A}^i$ which are not block diagonal cannot generally be created by a local operator, and therefore correspond to non-trivial domain walls (the flux domain walls).

Summarizing, the only non-trivial domain walls in symmetric (symmetry breaking) phases are charge (flux) domain walls, as claimed in Sec.~\ref{sec:phases}.




\section{Partially symmetry-broken phases} \label{sec:partialssb}

Throughout the main text, we focused on cases where the symmetry is either preserved or completely broken. When the symmetry is only partially broken, new phenomena can occur. Here we give one example in which SB phases with domain walls can only absorb SPT phases at the cost of creating charges of the preserved symmetry.

As a first example, consider a phase with breaks $Z_2\times Z_2$ symmetry down to $Z_2$. This can be represented by a Hamiltonian,
\begin{equation}
    H^u = \sum_i -Z^u_iZ^u_{i+1} - X^u_{i+\frac12}
\end{equation}
where the $Z_2$ symmetry acting on the integer (half-integer) sites is broken (preserved). Consider fusing this system with the cluster state,
\begin{equation}
H^l = \sum_i - Z^l_{i-\frac12}X^l_iZ^l_{i+\frac12} - Z^l_iX_{i+\frac12}Z^l_{i+1}.
\end{equation}
Now we apply the circuit depicted in Fig.~\ref{fig:SB+SPT}, except we only apply the $R(ZZ)$ and $R(ZXZ)$ gates that are centered on integer sites. This modifies the combined Hamiltonian to,
\begin{equation}
\begin{aligned}
    H = &\sum_i -Z^u_iZ^u_{i+1} - X^u_{i+\frac12} \\
    +&\sum_i - Z^u_iZ^l_i - Z^l_iX^l_{i+\frac12}Z^l_{i+1}.
\end{aligned}
\end{equation}
Using the fact that
\begin{equation} \label{eq:zxz}
    Z^l_iX^l_{i+\frac12}Z^l_{i+1} = X^l_{i+\frac12}(Z^u_iZ^l_i)(Z^u_iZ^u_{i+1})(Z^u_{i+1}Z^l_{i+1})
\end{equation}
we find that $Z^l_iX^l_{i+}Z^l_{i+1}\sim X^l_{i+1/2}$ where $\sim$ denotes equivalence in the ground space. Therefore, the Hamiltonian has the same ground space as,
\begin{equation}
\begin{aligned}
    H = &\sum_i -Z^u_iZ^u_{i+1} - X^u_{i+\frac12} \\
    +&\sum_i - Z^u_iZ^l_i - X^l_{i+\frac12}
\end{aligned}
\end{equation}
which describes a SB state on the integer sites and a trivial symmetric state on the half-integer sites. Therefore, even the partially symmetry-broken state is able to absorb the SPT phase. This is a reflection of the fact that the $Z_2\times Z_2$ SPT phase becomes trivial when only one of the $Z_2$ symmetries is enforced.

Now, suppose the initial system has a domain wall of the broken symmetry, indicated by setting $Z^u_{j}Z^u_{{j}+1}\sim-1$ at some site ${j}$. Applying the same circuit gives the same outcome, except Eq.~\ref{eq:zxz} now gives $Z^l_{j}X^l_{{j}+1/2}Z^l_{{j}+1}\sim - X^l_{{j}+1/2}$ Then, the ground state after applying the circuit satisfies $X^l_{{j}+1/2}=-1$, indicating the presence of a symmetry charge at site $j+1/2$. Therefore, fusing the partially symmetry-broken state with the SPT in the presence of a domain wall results in binding a charge of the unbroken symmetry to the domain wall.

The above calculation can be generalized to all abelian groups $G$. A general $G$-symmetric 1D phase with a defect can be labelled by a quadruple $(H,k;\omega,q)$. The 1D phase is specified by the the preserved symmetry subgroup $H \subset G$ and the SPT order $\omega\in H^2(H,U(1))$, while the defect is specified by the domain wall of the broken symmetry $k\in G/H$ and a charge of the unbroken symmetry $q\in H^1(H,U(1))$. Suppose we fuse a general SB state with a domain wall $(H,k;1,1)$ with a general SPT state $(G,1;\omega,1)$ (where $1$ represents the trivial element in all cases). Using cocycle states \cite{Chen2013}, the above calculation can be generalized to,
\begin{equation}
    (H,k;1,1) \times (G,1;\omega,1) \to (H,k,\omega|_H,\chi_{\omega}(k)).
\end{equation}

\noindent
Unpacking this, the result is that the $G$-SPT labelled by the cocycle $\omega$ is reduced to an $H$-SPT labelled by the restricted cocycle $\omega|_H\in H^2(H,U(1))$ defined as,
\begin{equation}
    \omega|_H(h_1,h_2)=\omega(h_1,h_2) \quad \forall h_1,h_2\in H.
\end{equation}
The $H$-charge created during the fusion is given by the slant product $\chi_\omega(k)$ which is defined as 
\begin{equation}
    \chi_{\omega}(k)(h) = \omega(k,h)\omega^{-1}(h,k) \quad \forall h\in H.
\end{equation}

\end{document}